\documentclass{article}
\usepackage[textwidth=6.9in,textheight=8.8in,top=0.85in]{geometry}
\usepackage{amsmath,amsfonts,amssymb,amsthm,mathtools}
\usepackage{fontspec}
\usepackage{booktabs,array,enumitem}
\usepackage{algorithm,algpseudocode}
\usepackage{titlesec}
\usepackage{authblk}

\usepackage[colorlinks=true,linkcolor=blue,citecolor=blue,urlcolor=blue]{hyperref}
\usepackage{tikz}
\usetikzlibrary{arrows.meta,positioning}
\titleformat{\subsubsection}[runin]
  {\normalfont\normalsize\bfseries}{\thesubsubsection}{1em}{}
\setlist{itemsep=0em,topsep=0.5em}
\newtheorem{theorem}{Theorem}[section]
\newtheorem{lemma}[theorem]{Lemma}
\newtheorem{proposition}[theorem]{Proposition}
\newtheorem{corollary}[theorem]{Corollary}

\newtheorem{definition}[theorem]{Definition}

\numberwithin{equation}{section}

\newcommand{\Tr}{\operatorname{Tr}}
\newcommand{\diag}{\operatorname{diag}}

\newcommand{\norm}[1]{\left\lVert#1\right\rVert}
\newcommand{\ket}[1]{\left|#1\right\rangle}
\newcommand{\bra}[1]{\left\langle#1\right|}
\newcommand{\F}{\mathcal F}

\hypersetup{linktoc=page}

\title{Efficient Classical Algorithms for Ground Energy Estimation of Fermions with Conjugate-Pair Orbital Interactions}

\author[1,2]{Ziyi Yang}
\author[1,2]{Yexin Zhang}
\author[1,2]{Yuexin Su}
\author[1,2]{Xinzhao Wang}
\author[3,4]{Xiaoyang Wang\thanks{Corresponding author. \texttt{xiaoyang.wang@riken.jp}}}
\author[1,2]{Tongyang~Li\thanks{Corresponding author. \texttt{tongyangli@pku.edu.cn}}}

\affil[1]{Center on Frontiers of Computing Studies, Peking University, Beijing 100871, China}
\affil[2]{School of Computer Science, Peking University, Beijing 100871, China}
\affil[3]{RIKEN Center for Interdisciplinary Theoretical and Mathematical Sciences (iTHEMS), Wako 351-0198, Japan}
\affil[4]{RIKEN Center for Computational Science (R-CCS), Kobe 650-0047, Japan}

\date{}

\begin{document}

\maketitle

\begin{abstract}
Estimating the ground energy of interacting fermions is central to
quantum many-body physics, and recent quantum algorithms provide
polynomial-time guarantees for some structured families.
Classical approaches commonly rely on quantum Monte Carlo, where the
sign problem is a major obstacle. However, even when a nonnegative expansion without sign problem
is available, rigorous guarantees for rapid mixing are still far from well understood.
In this paper, we give an efficient classical randomized algorithm for approximating the ground
energy of fermionic Hamiltonians with conjugate
hopping matrices and conjugate-paired orbital interactions,
which includes the attractive Hubbard models. This guarantee not only holds without a physical spectral gap or a weak-coupling
assumption, but also covers interactions beyond onsite density attractions. 
Technically, we first reduce ground-energy
estimation to partition-function estimation and then use a
discretization with controlled energy error to obtain a nonnegative
weighted sum of squared determinant moduli and then apply the Chen--Liu log-concave
counting theorem.
In addition, we obtain a sharper arithmetic bound by controlling how many independent sampling runs and matrix updates are needed, and using sparse matrices to reduce the cost of each update.
\end{abstract}

\section{Introduction}
\label{sec:introduction}
Computing the ground-state energy of interacting fermions is a central task in quantum many-body physics. It asks for the lowest energy compatible with the motion and interactions of the particles, and provides a basic starting point for understanding their low-energy behavior. The Hubbard model, as one of the simplest paradigmatic models of strongly correlated electrons, has played a central role in condensed matter physics, such as in understanding high-temperature superconductivity in cuprates and itinerant magnetism~\cite{Arovas_2022,RevModPhys.84.1383,PhysRevX.5.041041,Tasaki_1998,Hubbard1963}. This many-body system has Hilbert space dimension growing exponentially with the number of particle modes, including spins and spatial sites. Rigorous complexity results show the difficulty of solving the ground-state energy of the Hubbard model. For example, since the general two-local Hamiltonian problem is QMA-complete~\cite{KempeKitaevRegev2006}, the ground-energy estimation is QMA-complete even for the repulsive Hubbard model on a two-dimensional square lattice in the presence of site-dependent local magnetic fields~\cite{SchuchVerstraete2009}. These results motivate the search
for substantial families whose structure permits efficient algorithms
while retaining nontrivial many-body interactions.

On the one hand, recent advances in quantum computing have provided valuable insights into identifying such substantial families through the Lee--Yang theory~\cite{PhysRev.87.404,PhysRev.87.410}. For structured spin systems, bounds on the locations
of partition-function zeros imply field-induced spectral gaps and
polynomial-time quantum algorithms for ground-energy
estimation~\cite{RayuduTakahashi2026,BravyiGossetLiuWong2026}.
A weak field provides a quantitative gap while changing the ground
energy by a controlled amount, allowing the original energy to be
approximated without assuming that the original Hamiltonian is gapped.
A fermionic Lee--Yang theorem gives corresponding quantum guarantees
for a class that includes attractive Hubbard models with conjugate
spin-dependent hopping~\cite{RayuduMisawaZhaoTakahashi2026}.
However, these advances of quantum guarantees do not establish a quantum advantage, since the classical complexity of the same models remains a
separate question.

On the other hand, among classical algorithms, fermionic quantum Monte Carlo offers a natural approach to the ground-state problem. It evaluates the ground-state energy by calculating the partition function of the fermionic system. The partition function can be written as a weighted sum over configurations of auxiliary fields, and thus can be probabilistically calculated via Monte-Carlo sampling. Existence of negative or complex weights can lead to the sign problem~\cite{PhysRevB.41.9301}. To avoid this generally hard problem~\cite{PhysRevLett.94.170201}, algebraic structures, including conjugate pairing between two fermionic components and antiunitary symmetries, can yield nonnegative weights. 

In this work, we focus on the ground-state problem of conjugate-pair fermionic systems with orbital interactions. These systems include attractive Hubbard model studied in conventional BCS superconductivity with Cooper pairs~\cite{girvin2019modern,PhysRevResearch.5.023037,PhysRevB.85.014509}, and extend to more general conjugate-orbital attractions and Cooper-pair-hopping interactions on arbitrary hopping graphs~\cite{PhysRevB.33.1663,PhysRevB.48.2063,PhysRevB.59.6430}. Although the conjugate-pair fermionic systems with nonnegative weights are free from the sign problem, it does not by itself establish polynomial-time classical algorithms, since the mixing time of the Monte-Carlo Markov chain may still grow exponentially with system size. In summary, the following crucial question is raised:
\begin{quote}
\emph{Can we establish efficient classical algorithms for estimating the ground energy of fermionic models with conjugate-pair orbital interactions?}
\end{quote}
We answer this question affirmatively by giving polynomial guarantees on the sampling complexity and mixing time of the fermionic quantum Monte-Carlo Markov chain.

\subsection{Our results}
\label{intro:results}

Consider fermions with two components, labeled $\uparrow$ (spin-up) and $\downarrow$ (spin-down), each containing $n$ modes.
The full Hilbert space is the product of Fock spaces of the two components $\mathcal H_n\cong\F_n\otimes\F_n$, where the first and second factors correspond to $\uparrow$ and
$\downarrow$, respectively. For orbital vectors $u,v\in\mathbb C^n$, define the orbital creation and annihilation operators in $\F_n$ by
\begin{align}
 a^\dagger(u)=\sum_{i=1}^n u_i a_i^\dagger,
 \qquad a(v)=\sum_{i=1}^n\overline{v_i}a_i.
\end{align}
We study the conjugate-pair Hamiltonians with $s$ orbital interactions
\begin{align}\label{intro:model}
 H=cI+K-\frac12\sum_{\alpha=1}^s g_\alpha
       \left(A_\alpha\otimes\overline{A_\alpha}
           +A_\alpha^\dagger\otimes\overline{A_\alpha^\dagger}\right),
\end{align}        
where
\begin{align}
 K=d\Gamma(q)\otimes I+I\otimes d\Gamma(\overline q),
 \qquad A_\alpha=a^\dagger(u_\alpha)a(v_\alpha)
                  =d\Gamma(u_\alpha v_\alpha^\dagger),  \qquad d\Gamma(q)=\sum_{ij}q_{ij}a_i^\dagger a_j
\end{align}
satisfying $q=q^\dagger$,
$g_\alpha\geq0$, and bars denote conjugation in the occupation basis. The orbital vectors $u_{\alpha},v_{\alpha}$ may be complex, unnormalized, and
nonorthogonal, and the channels need not commute. The orbital transition operator $A_\alpha$ is defined for both $\uparrow$ and
$\downarrow$ components. It has a rank-one single-particle matrix
$u_\alpha v_\alpha^\dagger$. 
We note that this conjugate-pair Hamiltonians include important use cases. Taking $u_\alpha=v_\alpha$ to be the unit vector $e_i$ corresponding to the spatial site $i$ recovers onsite attractive Hubbard
interactions. $u_\alpha=e_i$, $v_\alpha=e_j$ describes the component pair hopping: the orbital interaction transfers a pair
from site $j$ to site $i$, together with the reverse transition, and the pair hopping graph is unrestricted. Section~\ref{p:hamiltonians} gives
the precise operator convention and the centered Hubbard example.

Our target is the smallest eigenvalue
\begin{equation}
 E_0(H)=\min\operatorname{spec}(H)
 \label{p:ground-energy}
\end{equation}
over the full space $\mathcal H_n$. Accuracy is measured by additive error $\varepsilon$ in the total energy. Floating-point arithmetic can introduce rounding errors to the overall additive error. Therefore, we use rational representations and bound the bit lengths of intermediate values in our complexity analysis. $c,g_\alpha,\varepsilon$ and the real and imaginary parts of $q,u_\alpha,v_\alpha$ are rational and encoded in
binary. A rational number is specified by an integer numerator and a
positive integer denominator, both written in binary. Let $b$
denote the total input encoding length. Define the rational strength
bound
\begin{equation}
 \lambda:=1+2\sum_{i,j=1}^n
        \bigl(|\operatorname{Re}q_{ij}|+|\operatorname{Im}q_{ij}|\bigr)
       +\sum_{\alpha=1}^s
           g_\alpha\norm{u_\alpha}_2^2\norm{v_\alpha}_2^2.
 \label{p:strength}
\end{equation}
Our main result for efficient ground energy estimation of fermions with conjugate-pair orbital interactions is:
\begin{theorem}[Main Theorem]
 \label{p:main}
 For the rational input above, there is a classical randomized algorithm that returns a
 rational number $\widehat E$ satisfying
 \begin{equation}
 \Pr\!\left[|\widehat E-E_0(H)|\leq\varepsilon\right]\geq\frac34.
 \label{p:energy-guarantee}
 \end{equation}
 Its running time in bit operations on every execution is
 $\operatorname{poly}(n,s,\lambda,\varepsilon^{-1},b)$.
 For $s\geq1$, after input preprocessing, constructing the matrices
 in the determinant representation and estimating its nonnegative sum require
 \begin{equation}
 \widetilde O\!\left([n(n+s)]^{13/4}
                 \lambda^{13/2}\varepsilon^{-13/2}\right)
 \label{intro:work}
 \end{equation}
additions, subtractions, multiplications, and divisions of rational numbers.\footnote{The notation $\widetilde O$ suppresses poly-logarithmic factors in the displayed parameters.} Preprocessing and the final logarithm evaluation have bit
costs polynomial in $n,s,\lambda,\varepsilon^{-1},b$ as well.
\end{theorem}

We point out that our result does not require physical spectral gap or weak-coupling assumption. In addition, if $s=0$, the model is quadratic and its discrete partition function is an exact determinant; no approximate-counting call is needed. 

\begin{figure}[H]
 \centering
 \begin{tikzpicture}[
   flow/.style={draw,rounded corners=2pt,align=center,
                inner sep=6pt,minimum height=1.05cm,font=\small},
   >={Latex[length=2mm]},semithick]
  \node[flow,text width=2.8cm] (energy) at (0,0)
       {Ground energy\\$E_0(H)$};
  \node[flow,text width=3.3cm] (thermal) at (5.1,0)
       {Thermal trace\\$Z_{H_0}(\beta)$};
  \node[flow,text width=3.5cm] (discrete) at (10.7,0)
       {Positive discrete trace\\$Z_{\mathrm{disc}}=\Tr S^M$};
  \node[flow,text width=9.1cm] (sum) at (5.35,-2.05)
       {Nonnegative determinant sum\\[2pt]
        $\displaystyle |\det C|^2
          \sum_{J}\gamma^J|\det F[J,J]|^2$};
  \node[flow,text width=5.0cm] (general) at (1.8,-4.05)
       {Section~\ref{sec:counting}: Chen--Liu counting\\
        Polynomial-time guarantee};
  \node[flow,text width=5.0cm] (specialized) at (8.9,-4.05)
       {Section~\ref{sec:fast}: Specialized counting\\
        Sharper arithmetic bound};
  \draw[->] (energy) -- node[above,align=center,font=\scriptsize]
       {thermal\\reduction} (thermal);
  \draw[->] (thermal) -- node[above,align=center,font=\scriptsize]
       {energy-error\\control} (discrete);
  \draw[->] (discrete.south) -- ++(0,-0.55) -|
       node[pos=0.25,above,font=\scriptsize] {exact expansion} (sum.north);
  \draw[->] (sum.south) -- (general.north);
  \draw[->] (sum.south) -- (specialized.north);
 \end{tikzpicture}
 \caption{Roadmap of our algorithms. Both algorithms use the same nonnegative representation: 
 Section~\ref{sec:counting} applies the general counting theorem;
 Section~\ref{sec:fast} refines the counting analysis and implementation.}
 \label{fig:proof-overview}
\end{figure}
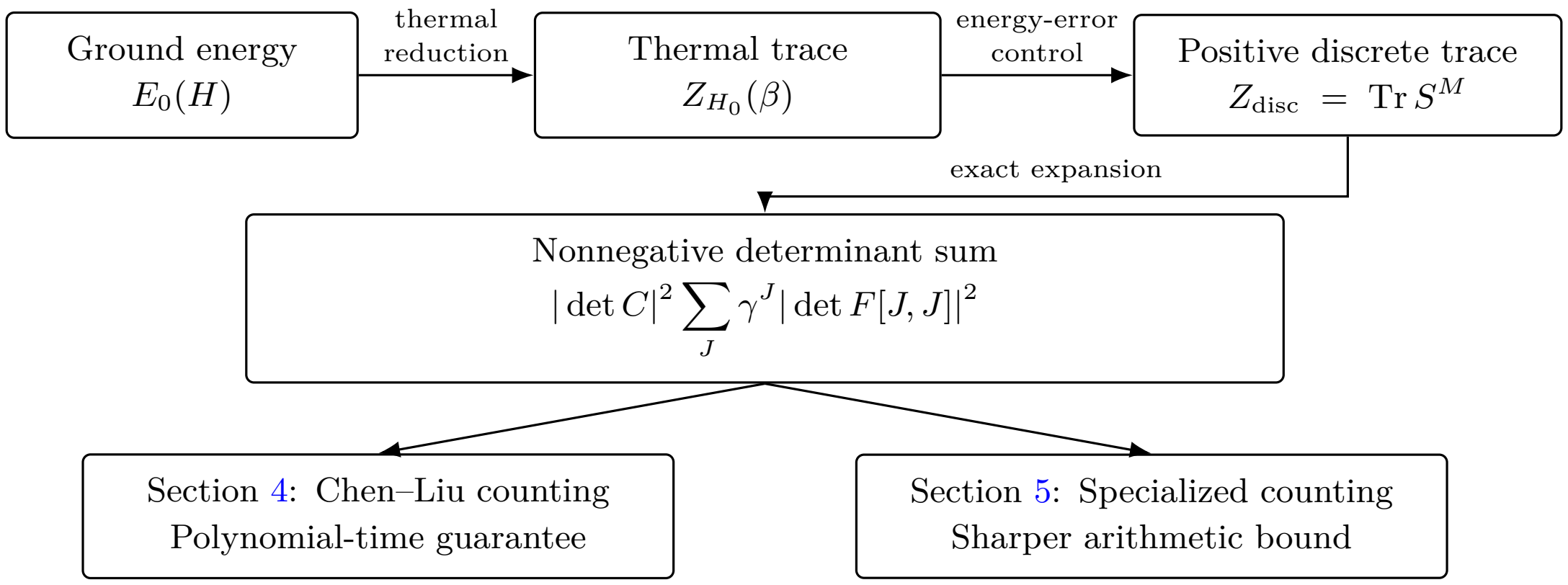

The two algorithms for Theorem~\ref{p:main} start from the same
thermal reduction: at inverse temperature
$\beta=\Theta(n/\varepsilon)$, a partition-function estimate with
constant relative accuracy suffices. Writing $H_0=H-cI$ to separate
the known scalar shift, we turn this estimation problem into a finite
counting problem by constructing a rational positive transfer with
controlled ground-energy error. Conjugate pairing then expresses its
discrete partition function as an exact nonnegative weighted sum of
squared determinant moduli. This sum is the common input to two
counting arguments: applying the Chen--Liu theorem in
Section~\ref{sec:counting} establishes the polynomial-time guarantee,
while a specialized sampling analysis and matrix implementation in
Section~\ref{sec:fast} give the sharper arithmetic bound for the same
full model class. Figure~\ref{fig:proof-overview} summarizes this roadmap.

\paragraph{From ground energy to a nonnegative sum (Section~\ref{sec:positive}).}
Since the full Fock space has dimension $4^n$, the quantity
$-\beta^{-1}\log\Tr e^{-\beta H}$ approximates $E_0(H)$ to within
$n\log4/\beta$. Thus choosing inverse
temperature $\beta=\Theta(n/\varepsilon)$ reduces the energy problem
to estimating a partition function with constant relative accuracy.
We make the construction rational by replacing the one-particle
exponential and interaction exponentials with first-order factors.
These factors form a product $X$; combining it with its adjoint gives
a positive step $S=XX^\dagger$. We bound the difference between
$H-cI$ and $-\delta^{-1}\log S$, where $\delta$ is the step size.
This controls the energy error without commuting the interaction
factors. The resulting discretized partition function satisfies
\begin{align}
 \left|E_0(H)-\left(c-\frac1\beta\log Z_{\mathrm{disc}}\right)\right|
     &\leq\frac\varepsilon2,
\end{align}    
where
\begin{align}
 Z_{\mathrm{disc}}=|\det C|^2\mathcal W,\qquad
 \mathcal W&=\sum_{J\subseteq[r]}\gamma^J|\det F[J,J]|^2.
 \label{eq:determinant}
\end{align}
Here $J$ selects from the $r$ ordered interaction events and
$\gamma^J=\prod_{j\in J}\gamma_j\geq0$. For each selection, the
rank-one insertions can be encoded by rank-one matrix updates, and
the Fock-space trace for one component becomes
$\det C\det F[J,J]$. The second component contributes its complex
conjugate, yielding the nonnegative squared modulus. The matrices
$C,F$ have polynomial dimension, so this representation can be
constructed without forming the many-body matrices.

\paragraph{Polynomial-time approximation by log-concave counting (Section~\ref{sec:counting}).}
To deal with the determinants in (\ref{eq:determinant}), 
we separate the representation of the partition function into two weight functions, one of which keeps the determinant value and the other records the choice of row
and column indices. Appending identity columns to $F$ gives
$\mathcal A=[I_r\mid F]$. Let $\mathcal U$ be its set of $2r$ column labels.
The function $\mu(S)=|\det\mathcal A_{:,S}|^2$ assigns an exact weight to every
$r$-column selection, including those that do not encode principal
minors. A second weight $\nu$ imposes the constraint: for each identity
column and its corresponding column of $F$, the complement of $S$
must select exactly one, and $\nu$ supplies the interaction coefficients.
Expanding along the selected identity columns recovers the original
sum as
\[
 \mathcal W=\sum_{S\subseteq\mathcal U}
                 \mu(S)\nu(\mathcal U\setminus S).
\]
Using one variable per column, the generating polynomial of $\mu$ is
$\det(\mathcal A\diag(x)\mathcal A^\dagger)$, while the independent
pair choices in $\nu$ give a product of linear forms. Both are log-concave,
every monomial has degree $r$, and each variable appears at most
linearly. We also give exact coefficient computations, sets with
known positive weight, and polynomial bounds on the logarithms of the ratios
between the largest and smallest positive coefficients. These checks
let us apply the counting theorem of Chen and Liu~\cite{ChenLiu2026},
which estimates $\mathcal W$ in randomized polynomial time and gives
the polynomial-time ground-energy guarantee in Theorem~\ref{p:main}.

\paragraph{A faster algorithm for the same model (Section~\ref{sec:fast}).}
Chen and Liu choose external fields by solving a convex program and
reduce counting to a sequence of sampling problems~\cite{ChenLiu2026}.
In our model, the constraint weight factors over pairs of columns.
We can therefore write down the external fields explicitly and sum out
the auxiliary set in their joint distribution, one pair at a time.
The remaining column set encodes a minor $F[I,J]$ with a factor
$\theta^{|I\triangle J|}$ penalizing unmatched row and column indices.
At $\theta=1$, the total weight $W_1$ is an explicit determinant.
At $\theta=0$, only principal minors contribute, giving the target
weight $W_0$. We estimate $W_0/W_1$ by annealed importance sampling.
At each positive temperature, we update only the column set. Averaging
the Chen--Liu exchange rates over the auxiliary set gives this chain
a spectral gap bound (Lemma~\ref{f:collapsed-chain}).
Conjugate pairing expresses $W_\theta$ as a positive operator trace
\eqref{f:source-trace}, whose growth controls the moments of
$|I\triangle J|$ even when the interaction channels overlap and do not
commute. These bounds guide the choice of cooling steps and control
the relative variance of the estimator. They also bound how often a
sampling step triggers a matrix update, since the probability of an
update depends on the number of unmatched indices. When an update is
needed, the sparse block structure of the rank-one insertions lets us
compute all replacement weights with one linear solve and update the
stored factorization after the exchange
(Lemmas~\ref{f:lift-scores} and~\ref{f:activation-cost}).
Writing $D=\Theta(n(n+s)\lambda^2\varepsilon^{-2})$ for the dimension
of the sparse matrix, we obtain the dominant arithmetic cost
\[
 \underbrace{\widetilde O(D^{7/4})}_{\text{number of matrix updates}}
 \;\times\;
 \underbrace{\widetilde O(D^{3/2})}_{\text{operations per update}}
 = \widetilde O(D^{13/4}).
\]
Substituting $D$ gives \eqref{intro:work}. For $s=O(n)$, as in the
onsite Hubbard model, this yields
$\widetilde O(n^{13/2}\lambda^{13/2}\varepsilon^{-13/2})$
rational arithmetic operations, with the preprocessing and bit costs
specified above.

\subsection{Related work}
\label{intro:related}

\paragraph{Quantum Monte Carlo and the sign problem.}
Two common approaches to fermionic quantum Monte Carlo differ in how
they express the partition function as a weighted sum or integral over
configurations to be sampled. In auxiliary-field methods, extra
variables are introduced so that, for each fixed choice of these
variables, the fermionic problem is noninteracting and its trace can
be evaluated using one-particle
matrices~\cite{BlankenbeclerScalapinoSugar1981,Hirsch1983}.
Interaction-expansion methods instead expand the partition function
in the interaction terms around a noninteracting reference; each
configuration records the inserted interaction terms and their
insertion times~\cite{RubtsovSavkinLichtenstein2005}. In the
number-conserving formulations relevant here, determinants enter the
weights in both constructions. The general sign problem has computational
complexity obstructions~\cite{TroyerWiese2005}, motivating sufficient
conditions for positive weights in particular representations.
Antiunitary-symmetry criteria~\cite{WuZhang2005}, Majorana reflection
positivity and time-reversal
criteria~\cite{WeiEtAl2016,LiJiangYao2016}, and semigroup
methods~\cite{Wei2024} provide several such conditions. These results
identify positive representations with nonnegative weights, while a polynomial mixing or counting
bound is an additional requirement. Our analysis exploits conjugate pairing in an interaction expansion and then establishes the polynomial counting
guarantees for that representation.

\paragraph{Classical algorithms for spin systems.}
Classical ferromagnetic Ising
approximation~\cite{JerrumSinclair1993} supports efficient algorithms for the
ferromagnetic transverse-field Ising model~\cite{Bravyi2015}, and
matching representations give polynomial-time simulation of
ferromagnetic XY systems~\cite{BravyiGosset2017}. Direct mixing
analyses also justify path-integral and operator-loop Monte Carlo
for particular families and parameter
regimes~\cite{CrossonHarrow2021,RayuduTakahashi2025}.

\paragraph{Classical algorithms for fermions.}
The quadratic structure of noninteracting fermionic systems permits efficient simulation~\cite{TerhalDiVincenzo2002}. For interacting systems, rigorous algorithms include Gaussian-mixture sampling at sufficiently
high temperature on bounded-degree interaction
graphs~\cite{RamkumarCaiTongJiang2025}, log-partition-function
approximation~\cite{ChenEtAl2025} and Gaussian-mixture
sampling~\cite{HanZhaoLiu2026} in suitable weak-interaction regimes
at fixed finite temperature, and ground-energy and equilibrium
algorithms for constant-size impurities coupled to noninteracting
baths~\cite{JiangEtAl2026,ArunachalamEtAl2026}. Our setup targets conjugate-pair orbital interactions and
zero-temperature ground-energy estimation with no restriction on the interaction strength.

\paragraph{Lee--Yang theory and quantum algorithms.}
The original Lee--Yang theorem concerns partition-function zeros of
classical lattice gases and Ising models~\cite{PhysRev.87.404,PhysRev.87.410}.
Extensions to quantum ferromagnets and more general spin systems
established analogous properties for noncommuting
interactions~\cite{Asano1970,SuzukiFisher1971}. Recent work turns
these properties into quantum ground-energy
algorithms~\cite{RayuduTakahashi2026,BravyiGossetLiuWong2026}, and
the fermionic theorem of Rayudu, Misawa, Zhao, and
Takahashi~\cite{RayuduMisawaZhaoTakahashi2026} includes attractive
Hubbard models with conjugate hopping and certain repulsive Hubbard
models after a particle--hole transformation. These quantum algorithms
apply to structured families that overlap with ours. Our classical
proof uses determinant counting and does not require a Lee--Yang
theorem for the full model class
\eqref{intro:model}.

\paragraph{Log-concave sampling and counting.}
Strong log-concavity and local spectral expansion underlie important
sampling and counting results for high-dimensional walks and matroid
bases~\cite{AnariEtAl2024}. Chen and Liu~\cite{ChenLiu2026} prove
an FPRAS for counting common bases of two matroids and a more general
counting theorem for complementary convolutions
\[
 Z(\mu,\nu)=\sum_{S\subseteq\mathcal U}
                    \mu(S)\nu(\mathcal U\setminus S),
\]
when the nonnegative weights have homogeneous, multiaffine,
log-concave generating polynomials, with the required oracle and
coefficient-range data. Their applications include the EPR model on
arbitrary graphs and Quantum Max-Cut on bipartite graphs. Our first
algorithm supplies an explicit fermionic instance of this theorem.
The refined analysis also uses their exchange inequality. It exploits
the determinant structure of our model to bound the variance of
importance sampling, the number of required matrix updates, and
the arithmetic cost of each update.

\subsection{Concurrent work}
\label{intro:concurrent}

In a concurrent work, Rayudu, Takahashi, Parekh, and
Zhao~\cite{RayuduEtAl2026} establish polynomial-time classical
algorithms for partition functions, free energies, and ground energies
of the \emph{attractive Kramers class}:
\begin{equation}
 H_{\mathrm K}=d\Gamma(h)-\sum_\alpha U_\alpha
                    n(v_\alpha)n(\Theta v_\alpha),
 \qquad U_\alpha\geq0,
 \label{intro:concurrent-class}
\end{equation}
where $h=h^\dagger$, $\Theta h\Theta^{-1}=h$, and $\Theta$ is an
antiunitary (a conjugate-linear isometry) satisfying $\Theta^2=-I$.
Here $n(v)=a^\dagger(v)a(v)$ is the occupation operator of a unit
orbital $v$ in the full one-particle space.

Our model fixes $h=q\oplus\overline q$ but allows the transitions
$A_\alpha=a^\dagger(u_\alpha)a(v_\alpha)$ in \eqref{intro:model},
including $u_\alpha\ne v_\alpha$. These interactions include pair
hopping outside the density-interaction form
\eqref{intro:concurrent-class}. Both our model and the model in~\cite{RayuduEtAl2026} contain the attractive
Hubbard model with conjugate hopping, but neither contains the other;
Appendix~\ref{sec:comparison-examples} gives concrete examples explaining the relationship between the two models, especially models that we can cover but they cannot.

In Section~\ref{sec:fast}, we use the structure of our determinant
representation to improve the polynomial bound obtained by directly
applying the general Chen--Liu counting algorithm~\cite{ChenLiu2026}.
For $s=O(n)$, we obtain
$\widetilde O(n^{13/2}\lambda^{13/2}\varepsilon^{-13/2})$
rational arithmetic operations, with preprocessing and bit costs as
specified in Theorem~\ref{p:main}. In contrast, Ref.~\cite{RayuduEtAl2026} prove polynomial running time using a worm chain and the Chen--Liu mixing framework, but their main theorem does not specify the polynomial degree.

In addition, there is a polynomial-time algorithm for estimating the
ground energy of a larger class containing both our model and the
model in~\cite{RayuduEtAl2026}. Specifically, set
$\Delta(v)=a(\Theta v)a(v)$ and consider
\begin{equation}
 \begin{aligned}
 H_{\mathrm{ext}}={}&cI+d\Gamma(h)
 -\frac12\sum_\alpha g_\alpha
 \bigl[\Delta^\dagger(u_\alpha)\Delta(v_\alpha)
       +\Delta^\dagger(v_\alpha)\Delta(u_\alpha)\bigr],\\
 &h=h^\dagger,\qquad \Theta h\Theta^{-1}=h,\qquad g_\alpha\geq0.
 \end{aligned}
 \label{intro:common-extension}
\end{equation}
Taking $u_\alpha=v_\alpha$ recovers \eqref{intro:concurrent-class}.
For $\Theta(x,y)=(-\overline y,\overline x)$, taking
$h=q\oplus\overline q$ and all endpoints in the first component recovers
\eqref{intro:model}. For rational input in a supplied
Kramers-adapted basis, the running time is polynomial in the input
length, strength bound, and inverse accuracy. The proof combines a
rank-two determinant expansion with the quaternionic minor-polynomial
identity of~\cite[Lemma~8]{RayuduEtAl2026} and the Chen--Liu counting
theorem; see Appendix~\ref{sec:common-extension}.

\subsection{Organization}
We first introduce the notation and standard identities in
Section~\ref{sec:prelim}. Then we reduce ground-energy approximation to a
nonnegative determinant sum in Section~\ref{sec:positive}. This sum
is the starting point for two counting arguments: applying the
Chen--Liu theorem in Section~\ref{sec:counting} establishes the
polynomial-time guarantee, while a counting algorithm tailored to
this sum in Section~\ref{sec:fast} yields the sharper arithmetic
bound for the same model class. The appendices provide the remaining proof
details, examples distinguishing the classes, and the polynomial-time
algorithm for that larger class.

\section{Preliminaries}
\label{sec:prelim}

We begin by describing fermionic Fock space and its creation,
annihilation, and occupation operators in
Section~\ref{p:fermionic-notation}. Building on this notation, we
introduce the additive and multiplicative second-quantization maps,
$d\Gamma$ and $\Gamma$, in
Section~\ref{p:second-quantization-subsection}. These maps lift
one-particle matrices to operators on Fock space. We record their
basic identities, which will be used in Section~\ref{sec:positive}
to convert many-body traces into matrix determinants. Finally, using
these maps and a two-component Fock space, we formally define our
conjugate-pair rank-one Hamiltonians in
Section~\ref{p:hamiltonians}.

\subsection{Fock space and occupation operators}
\label{p:fermionic-notation}

Write $[n]=\{1,\ldots,n\}$. For a matrix $X$, let $\overline X$ denote
entrywise complex conjugation and $X^\dagger$ its adjoint. We use
$\norm{X}$ for the operator norm and
$\norm{X}_1=\Tr\sqrt{X^\dagger X}$ for the trace norm. For ordered
index sets $J,K$, write $X[J,K]$ for the corresponding submatrix.
The empty determinant is $1$. Identity operators are denoted by $I$,
with their dimensions indicated when needed. The notation $X\succ0$
means that $X$ is Hermitian positive definite, and $X\succeq Y$ means
that $X-Y$ is Hermitian positive semidefinite. All logarithms are natural
unless a base is specified.

We first describe a single fermionic species, also called a
\emph{component}, with $n\geq1$ modes. A mode is a one-particle basis
state; depending on the model, it may correspond to a site or to a
chosen orbital. The occupation-number representation of its Fock
space is
\begin{equation}
 \F_n=\operatorname{span}\{\ket{J}:J\subseteq[n]\},
 \qquad \dim\F_n=2^n.
 \label{p:fock-space}
\end{equation}
The vectors $\ket{J}$ form an orthonormal basis. The set $J$ records
the occupied modes, each of which can be occupied at most once, and
$\ket{\varnothing}$ is the vacuum. We write
$\F_n^{(k)}=\operatorname{span}\{\ket{J}:|J|=k\}$ for the
$k$-particle sector. Thus the full Fock space includes every particle
number from $0$ to $n$.

The annihilation operator $a_i$ removes a particle in mode $i$, with
the sign determined by its position in the ordered occupation basis:
\begin{equation}
 a_i\ket{J}=
 \begin{cases}
 (-1)^{|J\cap[i-1]|}\ket{J\setminus\{i\}},&i\in J,\\
 0,&i\notin J,
 \end{cases}
 \qquad [0]=\varnothing.
 \label{p:annihilation}
\end{equation}
Its adjoint $a_i^\dagger$ creates a particle in mode $i$, and is zero
on a state in which that mode is already occupied. These operators
satisfy the canonical anticommutation relations
\begin{equation}
 \{a_i,a_j^\dagger\}=\delta_{ij}I,
 \qquad \{a_i,a_j\}=\{a_i^\dagger,a_j^\dagger\}=0,
 \label{p:car}
\end{equation}
where $\{A,B\}=AB+BA$. In this convention
$\ket{J}=a_{j_1}^\dagger\cdots a_{j_k}^\dagger\ket{\varnothing}$
for $j_1<\cdots<j_k$, and the matrices of $a_i$ are real. An operator $a_i^\dagger a_j$ first removes a
particle and then creates one, so it preserves total particle number.
A linear combination of these operators is called a
\emph{number-conserving quadratic operator}. The occupation operator $n_i=a_i^\dagger a_i$ is the diagonal
projection satisfying $n_i\ket{J}=\mathbf1_{\{i\in J\}}\ket{J}$.
The total particle-number operator is $\mathsf N=\sum_i n_i$.

For $u\in\mathbb C^n$, write
\begin{equation}
 a^\dagger(u)=\sum_i u_i a_i^\dagger,
 \qquad a(u)=\sum_i\overline{u_i}a_i.
 \label{p:orbital-operators}
\end{equation}
If $\norm{u}_2=1$, then $a^\dagger(u)$ creates a particle in orbital
$u$ and $a^\dagger(u)a(u)$ measures its occupation. The anticommutation relations give
$\norm{a(u)}=\norm{a^\dagger(u)}=\norm{u}_2$, and consequently
\begin{equation}
 \norm{a^\dagger(u)a(v)}\leq\norm{u}_2\norm{v}_2.
 \label{p:bilinear-norm}
\end{equation}
Indeed, $a(u)a^\dagger(u)+a^\dagger(u)a(u)=\norm{u}_2^2I$
gives the upper bound, and applying $a^\dagger(u)$ to the vacuum gives
equality for its norm. 

\subsection{Second quantization}
\label{p:second-quantization-subsection}

A one-particle matrix $X$ acts on $\mathbb C^n$. We will use two ways of
lifting such a matrix to the whole Fock space, which are called additive and multiplicative second quantization maps.

\begin{definition}[Additive and multiplicative second quantization]
 \label{p:second-quantization}
 For $X\in\mathbb C^{n\times n}$, define
 \begin{equation}
 \begin{aligned}
 d\Gamma(X)&=\sum_{i,j=1}^nX_{ij}a_i^\dagger a_j,\\
 \Gamma(X)\ket{K}
 &=\left[\prod_{\ell=1}^k
       \left(\sum_{i=1}^nX_{ij_\ell}a_i^\dagger\right)\right]
          \ket{\varnothing},
 \qquad K=\{j_1<\cdots<j_k\}.
 \end{aligned}
 \label{p:second-quantization-formulas}
 \end{equation}
 The product is ordered from left to right by increasing $\ell$.
 The vacuum block of $\Gamma(X)$ is the scalar $1$, whereas
 $d\Gamma(X)\ket{\varnothing}=0$.
\end{definition}

Both constructions preserve particle number and reduce to $X$ on the
one-particle sector. In an occupation state, $\Gamma(X)$ replaces
every creation operator $a_j^\dagger$ by
$\sum_iX_{ij}a_i^\dagger$, keeping their original order.
The operator $d\Gamma(X)$ sums the terms obtained by making this
replacement at one occupied mode at a time. Neither definition
requires $X$ to be invertible or Hermitian.

\begin{lemma}[Second-quantization identities]
 \label{p:second-quantization-identities}
 For arbitrary $X,Y\in\mathbb C^{n\times n}$,
 \begin{equation}
 \bra{J}\Gamma(X)\ket{K}=
 \begin{cases}
 \det X[J,K],&|J|=|K|,\\
 0,&|J|\ne|K|.
 \end{cases}
 \label{p:minor-amplitudes}
 \end{equation}
 Moreover,
 \begin{equation}
 \Gamma(XY)=\Gamma(X)\Gamma(Y),\qquad
 \Gamma(e^X)=e^{d\Gamma(X)},\qquad
 \Tr\Gamma(X)=\det(I_n+X).
 \label{p:lift-identities}
 \end{equation}
 The trace in the last identity is over $\F_n$, whereas its
 determinant is of an $n\times n$ matrix. The lifts also satisfy
 \begin{equation}
 \Gamma(X^\dagger)=\Gamma(X)^\dagger,
 \qquad \Gamma(\overline X)=\overline{\Gamma(X)}.
 \label{p:lift-adjoints}
 \end{equation}
\end{lemma}
\begin{proof}
 Expanding the product in
 \eqref{p:second-quantization-formulas}, the anticommutation relations
 assign exactly the permutation signs in the determinant. This proves
 \eqref{p:minor-amplitudes}; see also
 \cite[Section~III]{TerhalDiVincenzo2002}. Cauchy--Binet then proves
 the multiplication identity on every particle-number sector.

 To prove the exponential identity, differentiate
 $\Gamma(e^{tX})$ at $t=0$. On a $k$-particle vector, this replaces
 one occupied orbital at a time by its image under $X$, giving
 $d\Gamma(X)$. The multiplication identity implies
 $\frac{d}{dt}\Gamma(e^{tX})=d\Gamma(X)\Gamma(e^{tX})$, with initial
 value $\Gamma(I_n)=I$, and hence
 $\Gamma(e^{tX})=e^{t\,d\Gamma(X)}$.
 Finally, summing the diagonal matrix elements gives
 \[
 \Tr\Gamma(X)=\sum_{J\subseteq[n]}\det X[J,J]
              =\det(I_n+X)
 \]
 by the principal-minor expansion. The adjoint and conjugation
 identities follow from the corresponding identities for minors.
\end{proof}

The exponential identity connects the additive and multiplicative second quantization maps, while the trace identity reduces a
Fock-space trace to a determinant~\cite{Klich2002}. We next record
the spectral and norm bounds that will control the approximation
error in Section~\ref{sec:positive}.

\begin{lemma}[Spectrum and norm]
 \label{p:spectrum}
 If $X$ is Hermitian with eigenvalues $\xi_1,\ldots,\xi_n$, the
 eigenvalues of $d\Gamma(X)$ are the subset sums
 $\sum_{j\in J}\xi_j$, indexed by $J\subseteq[n]$. Consequently,
 \begin{equation}
 \norm{d\Gamma(X)}\leq\norm{X}_1.
 \label{p:additive-norm}
 \end{equation}
 The eigenvalues of $\Gamma(X)$ are the subset products
 $\prod_{j\in J}\xi_j$, including the empty product $1$.
 In particular, $X\succ0$ implies $\Gamma(X)\succ0$.
\end{lemma}
\begin{proof}
 Write $X=UDU^\dagger$ with $U$ unitary and
 $D=\diag(\xi_1,\ldots,\xi_n)$. The preceding lemma shows that
 $\Gamma(U)$ is unitary and reduces $\Gamma(X)$ to $\Gamma(D)$.
 Differentiating the same identity for $e^{tX}$ reduces $d\Gamma(X)$
 to $d\Gamma(D)=\sum_j\xi_jn_j$. In the occupation basis these
 diagonal operators have the stated entries. Their subset sums have
 absolute value at most $\sum_j|\xi_j|=\norm{X}_1$.
\end{proof}

These bounds will control the quadratic part of the Hamiltonian. In particular, the positivity of a one-particle matrix is preserved by
$\Gamma$. To treat the interaction factors, we also need an exact
relation between rank-one matrix updates and operators on Fock
space, which is given next.

\begin{lemma}[A rank-one update lifts to a linear insertion]
 \label{p:rank-one-lift}
 For $u,v\in\mathbb C^n$ and $z\in\mathbb C$,
 \begin{equation}
 d\Gamma(uv^\dagger)=a^\dagger(u)a(v),\qquad
 \Gamma(I_n+zuv^\dagger)=I+z\,a^\dagger(u)a(v).
 \label{p:rank-one-lift-formula}
 \end{equation}
\end{lemma}
\begin{proof}
 The first identity follows by expanding the coefficients. For the
 second, let $K=\{j_1<\cdots<j_k\}$. The definition of $\Gamma$ gives
 \[
  \Gamma(I_n+zuv^\dagger)\ket{K}
  =\left[\prod_{\ell=1}^k
     \bigl(a_{j_\ell}^\dagger+
             z\overline{v_{j_\ell}}a^\dagger(u)\bigr)\right]
       \ket{\varnothing}.
 \]
 The anticommutation relations imply $(a^\dagger(u))^2=0$ and
 $\{a^\dagger(u),a_i^\dagger\}=0$, so every term containing two or
 more factors of $z$ vanishes. Moving the single $a^\dagger(u)$ to
 the left shows that the coefficient of $z$ is
 \[
  \sum_{\ell=1}^k(-1)^{\ell-1}\overline{v_{j_\ell}}\,
        a^\dagger(u)\ket{K\setminus\{j_\ell\}}
     =a^\dagger(u)a(v)\ket{K},
 \]
 by \eqref{p:annihilation}. The constant term is $\ket{K}$, proving
 the identity on every occupation state, including the vacuum.
\end{proof}

The rank-one identity \eqref{p:rank-one-lift-formula} will let us
encode the interaction insertions in Section~\ref{sec:positive}
by matrix updates.

\subsection{Conjugate-pair rank-one Hamiltonians}
\label{p:hamiltonians}

We now use two components, labeled $\uparrow$ and $\downarrow$, each
with $n$ modes. Ordering all up-component modes before all down-component modes identifies
the full space with
\begin{equation}
 \mathcal H_n\cong\F_n\otimes\F_n,
 \qquad \dim\mathcal H_n=4^n.
 \label{p:two-component-space}
\end{equation}
To represent physical creation and annihilation operators, let
$P\ket{J}=(-1)^{|J|}\ket{J}$ and set
\begin{equation}
 c_{i\uparrow}=a_i\otimes I,
 \qquad c_{i\downarrow}=P\otimes a_i.
 \label{p:parity-convention}
\end{equation}
The parity factor ensures anticommutation between different components.
It cancels in quadratic terms within the down component:
\begin{equation}
 c_{i\uparrow}^\dagger c_{j\uparrow}
   =a_i^\dagger a_j\otimes I,
 \qquad
 c_{i\downarrow}^\dagger c_{j\downarrow}
   =I\otimes a_i^\dagger a_j.
 \label{p:bilinear-tensor-convention}
\end{equation}
Thus products of these number-conserving quadratic operators have the
ordinary tensor representation below.

\begin{definition}[Conjugate-pair Hamiltonian]
 \label{p:model-definition}
 Let $q=q^\dagger\in\mathbb C^{n\times n}$, $c\in\mathbb R$,
 $g_\alpha\geq0$, and $u_\alpha,v_\alpha\in\mathbb C^n$ for
 $\alpha\in[s]$. Define operators on $\mathcal H_n$ by
 \begin{equation}
 \begin{aligned}
 A_\alpha&=a^\dagger(u_\alpha)a(v_\alpha)
              =d\Gamma(u_\alpha v_\alpha^\dagger),\qquad
 B_\alpha=A_\alpha\otimes\overline{A_\alpha},\\
 K&=d\Gamma(q)\otimes I+I\otimes d\Gamma(\overline q),\\
 H&=cI+K-\frac12\sum_{\alpha=1}^s g_\alpha
                         (B_\alpha+B_\alpha^\dagger).
 \end{aligned}
 \label{p:model}
 \end{equation}
 Conjugation is taken in the occupation basis. We call each pair
 $(u_\alpha,v_\alpha)$ an interaction channel.
\end{definition}

The one-particle matrices of the two components are $q$ and
$\overline q$. We require $q$ to be Hermitian, but impose no realness,
positivity, bipartiteness, or restriction on its hopping graph. Its
diagonal is real, so the diagonal one-particle potentials are common
to the two components. The word \emph{rank-one} refers to the
one-particle matrix $u_\alpha v_\alpha^\dagger$, rather than to
$A_\alpha$ as an operator on Fock space. The channels may overlap
and need not commute. We allow $u_\alpha\ne v_\alpha$, so
$A_\alpha$ and $B_\alpha$ may individually be non-Hermitian; including
$B_\alpha^\dagger$ makes $H$ Hermitian. The endpoint vectors $u_{\alpha}$ and $v_{\alpha}$ need not
be normalized or orthogonal.

For example, $u_\alpha=v_\alpha=e_i$ gives $A_\alpha=n_i$ and
$B_\alpha=n_i\otimes n_i$. Consequently, the attractive
Hubbard Hamiltonian
\begin{equation}
 \begin{aligned}
 H_{\mathrm{Hub}}(T,U)
 ={}&\sum_{i,j}
 \left(T_{ij}c_{i\uparrow}^\dagger c_{j\uparrow}
 +\overline{T_{ij}}c_{i\downarrow}^\dagger c_{j\downarrow}\right)\\
 &-\sum_i U_i
   \left(n_{i\uparrow}-\frac12I\right)
   \left(n_{i\downarrow}-\frac12I\right),
 \qquad T=T^\dagger,\quad U_i\geq0,
 \end{aligned}
 \label{p:hubbard-example}
\end{equation}
is obtained by taking $g_i=U_i$, $u_i=v_i=e_i$,
$q=T+\frac12\diag(U)$, and $c=-\frac14\sum_iU_i$.
For a general normalized $u=v$, the same construction gives attraction
between orbital $u$ in one component and its conjugate in the other.
For $u=e_i,v=e_j$ with $i\ne j$, it instead moves one particle in
each component from mode $j$ to mode $i$, together with the reverse
transition supplied by the adjoint. These transition channels explain
why the class extends beyond onsite density interactions.

\section{From ground energy to a nonnegative determinantal expansion}
\label{sec:positive}

In Section~\ref{p:thermal-subsection}, we relate a
partition-function estimate to the energy accuracy in
Theorem~\ref{p:main} and choosing the temperature and time step.
With these parameters fixed, Section~\ref{d:transfer-section}
constructs a rational positive transfer and bounds its effect on the
ground energy. We then track the ordered interaction insertions in
Section~\ref{d:events-section}; a cyclic determinant identity in
Section~\ref{d:cyclic-section} converts their expansion into a
nonnegative determinant sum.

\subsection{Thermal reduction and parameter choice}
\label{p:task}
\label{p:thermal-subsection}

We use the rational input convention and strength parameter $\lambda$
from Section~\ref{intro:results}.
The parameter $\lambda$ bounds the nonconstant part of the Hamiltonian:
\begin{equation}
 \norm{H-cI}\leq2\norm{q}_1
       +\sum_\alpha g_\alpha
                \norm{u_\alpha}_2^2\norm{v_\alpha}_2^2
       \leq\lambda-1.
\end{equation}
Here we used Lemma~\ref{p:spectrum}, \eqref{p:bilinear-norm}, and
$\norm q_1\leq\sum_{i,j}|q_{ij}|$.

For a Hermitian operator $A$ and inverse temperature $\beta>0$, write
$Z_A(\beta)=\Tr e^{-\beta A}$. We use the following elementary bound.

\begin{lemma}[Thermal approximation]
 \label{p:thermal}
 Let $A$ be Hermitian on a $d$-dimensional Hilbert space. Then
 \begin{equation}
 E_0(A)-\frac{\log d}{\beta}
 \leq-\frac1\beta\log Z_A(\beta)\leq E_0(A).
 \label{p:thermal-bound}
 \end{equation}
\end{lemma}
\begin{proof}
 Every eigenvalue of $A$ is at least $E_0(A)$, and at least one equals
 it. Therefore
 $e^{-\beta E_0(A)}\leq Z_A(\beta)\leq d e^{-\beta E_0(A)}$.
 Taking logarithms proves the claim.
\end{proof}

Since $\dim\mathcal H_n=4^n$, choosing
$\beta=\Theta(n/\varepsilon)$ bounds the thermal error by a constant
multiple of $\varepsilon$. Only constant relative accuracy in the
partition function is required: if
$(3/4)Z_A(\beta)\leq\widehat Z\leq(5/4)Z_A(\beta)$, then
\begin{equation}
 \left|-\frac1\beta\log\widehat Z
             +\frac1\beta\log Z_A(\beta)\right|
 \leq\frac{\log(4/3)}{\beta}.
 \label{p:relative-count-to-energy}
\end{equation}
Neither bound depends on the spectral gap of $A$.

For the discrete-time reduction, choose the rational parameters
\begin{equation}
 \beta=\frac{16n}{\varepsilon},\qquad
 M=\left\lceil\frac{200\beta\lambda^2}{\varepsilon}\right\rceil,
 \qquad\delta=\frac\beta M.
 \label{c:parameters}
\end{equation}
Then $\delta\lambda^2\leq\varepsilon/200$ and
$\delta\lambda\leq1/200$, so all transfer estimates below apply.

With these parameters fixed, the next subsection constructs a
rational positive transfer whose ground-energy error fits the
remaining budget. Its trace will be evaluated through matrices of
polynomial dimension and determinant data.

\subsection{A rational positive transfer operator}
\label{d:transfer-section}

We construct a rational discrete-time approximation to the partition
function of the Hamiltonian in Definition~\ref{p:model-definition}. Expanding its
interaction factors gives a sum of squared moduli of determinants.
Conjugate pairing is used at this step: the two components contribute
complex-conjugate traces even when the channels overlap or fail to
commute.

Write $H_0=H-cI$ and retain the strength parameter $\lambda$ from
\eqref{p:strength}. The norm bounds in Section~\ref{sec:prelim} give
\begin{equation}
 \norm K+\sum_{\alpha=1}^s g_\alpha\norm{B_\alpha}
 \leq\lambda-1,\qquad 2\norm q_1\leq\lambda-1.
 \label{d:strength-bound}
\end{equation}
Here $\norm q_1$ denotes the trace norm of $q$, the sum of its singular
values. Channels with zero strength or a zero endpoint are discarded;
throughout the remaining sections $s$ counts the retained channels.

Let $\delta>0$ be rational, with $\delta\lambda\leq1/64$, and set
\begin{equation}
 \begin{split}
 R&=I_n-\frac\delta2q,\qquad
 G=\Gamma(R)\otimes\Gamma(\bar R),\qquad
 \gamma_\alpha=\frac{\delta g_\alpha}{2},\\
 P&=\prod_{\alpha=1}^s(I+\gamma_\alpha B_\alpha),\qquad
 X=GP,\qquad S=XX^\dagger.
 \end{split}
 \label{d:transfer}
\end{equation}
Products are ordered from left to right in increasing $\alpha$.
For a positive integer $M$ and $\beta=M\delta$, define
\begin{equation}
 Z_{\mathrm{disc}}=\Tr S^M.
 \label{d:discrete-partition}
\end{equation}
These many-body operators are used to analyze the construction. The
algorithm will evaluate $Z_{\mathrm{disc}}$ through matrices whose
dimensions are polynomial in $n,s,M$.

The construction follows the first-order terms of the desired
exponential. Lemma~\ref{p:second-quantization-identities} converts
$\Gamma(e^{-\delta q/2})$ into $e^{-\delta d\Gamma(q)/2}$, and $R$
replaces $e^{-\delta q/2}$ by its first-order Taylor approximation.
Likewise, $I+\gamma_\alpha B_\alpha$ retains the linear term of the
corresponding interaction exponential. Consequently
$X=I+\delta[-K/2+\frac12\sum_\alpha g_\alpha B_\alpha]+O(\delta^2)$,
so $XX^\dagger=I-\delta H_0+O(\delta^2)$. Using $XX^\dagger$ also
ensures a Hermitian positive transfer step without requiring the
individual $B_\alpha$ to be Hermitian.

\begin{proposition}[Effective-Hamiltonian error]
 \label{d:transfer-error}
 If $\delta\lambda\leq1/64$, then $R\succ0$, $S\succ0$, and
 \begin{equation}
 H_{\mathrm{eff}}=-\delta^{-1}\log S,\qquad
 Z_{\mathrm{disc}}=\Tr e^{-\beta H_{\mathrm{eff}}},\qquad
 \norm{H_{\mathrm{eff}}-H_0}\leq10\delta\lambda^2.
 \label{d:effective-Hamiltonian}
 \end{equation}
\end{proposition}
The logarithm is defined by diagonalizing the positive definite
Hermitian matrix $S$ and taking the real logarithm of each eigenvalue.
Thus $H_{\mathrm{eff}}$ is Hermitian. The proof in
Appendix~\ref{a:transfer-proof} bounds the constant and linear terms
of the ordered product and then applies the convergent logarithmic
series. It uses no commutation assumption between channels.

\subsection{Ordered interaction events and a single-component polynomial}
\label{d:events-section}

There are $r=2sM$ interaction events in the expansion of $S^M$.
Label macrosteps by $m\in\{0,\ldots,M-1\}$. Within each macrostep,
list the forward channels $1,\ldots,s$, followed by the adjoint
channels $s,\ldots,1$. Flatten this ordered list into $j\in[r]$.
At a forward event for channel $\alpha$, put
$(u_j,v_j,\gamma_j)=(u_\alpha,v_\alpha,\gamma_\alpha)$; at its
adjoint event, put
$(u_j,v_j,\gamma_j)=(v_\alpha,u_\alpha,\gamma_\alpha)$.
We write $m(j)$ for the macrostep containing event $j$ and
$A_j=a^\dagger(u_j)a(v_j)$.

Since $S=GPP^\dagger G$ and $G$ is Hermitian, trace cyclicity gives
\[
 \Tr S^M=\Tr(G^2PP^\dagger)^M.
\]
Thus the free factors between consecutive macrosteps are $G^2$.
Introduce one formal variable $z_j$ for each event and define
\begin{equation}
 \begin{split}
 f(z)&=\Tr_{\F_n}\prod_{m=0}^{M-1}
       \left[\Gamma(R^2)
          \prod_{j:\,m(j)=m}(I+z_jA_j)\right],\\
 f(z)&=\sum_{J\subseteq[r]}a_Jz^J,
 \qquad z^J=\prod_{j\in J}z_j.
 \end{split}
 \label{d:source-polynomial}
\end{equation}
Each variable occurs in one linear factor, so $f$ is multiaffine:
its degree in each variable is at most one. The coefficients $a_J$
are ordered single-component traces and need not be real.
The rank-one identity in Lemma~\ref{p:rank-one-lift} and the
second-quantization identities give the smaller determinant
\begin{equation}
 f(z)=\det\!\left(I_n+
       \prod_{m=0}^{M-1}
        \left[R^2\prod_{j:\,m(j)=m}
                 (I_n+z_ju_jv_j^\dagger)\right]\right).
 \label{d:source-product-determinant}
\end{equation}
The products in this formula retain the event order. Reordering
overlapping channels would generally change the polynomial.

\subsection{A cyclic linearization}
\label{d:cyclic-section}

We next express all coefficients of $f$ as principal minors of a
single rational matrix. The following form also allows us to regroup
events later without changing the determinant.

\begin{lemma}[Cyclic source determinant]
 \label{d:cyclic-source}
 Partition the ordered events into $k$ consecutive nonempty chunks,
 with each macrostep partitioned separately. Let $a(j)\in[k]$ be
 the chunk containing event $j$. Set $T_a=R^2$ if chunk $a$ is the
 first chunk of a macrostep, and $T_a=I_n$ otherwise.
 For $k\geq2$, form the $nk\times nk$ block matrix
 \begin{equation}
 C_k=I_{nk}+\mathcal B_k,\qquad
 (\mathcal B_k)_{a,a+1}=-T_{a+1}\ (a<k),\qquad
 (\mathcal B_k)_{k,1}=T_1,
 \label{d:cyclic-matrix}
 \end{equation}
 with all other blocks zero; for $k=1$, set $C_1=I_n+T_1$.
 Let $E_a:\mathbb C^n\to\mathbb C^{nk}$ embed into block $a$, and
 define
 \begin{equation}
 \widehat U_{:,j}=E_{a(j)}u_j,\qquad
 \widehat V_{:,j}=E_{a(j)}v_j,\qquad
 L_{j\ell}=\mathbf1_{\{a(j)=a(\ell),\,\ell\leq j\}}
                      v_j^\dagger u_\ell.
 \label{d:cyclic-endpoints}
 \end{equation}
 Then $C_k$ is invertible, and the same matrix $F\in\mathbb C^{r\times r}$
 is obtained from every such partition by
 \begin{equation}
 \begin{split}
 F&=L-\widehat V^\dagger C_k^{-1}\widehat U,\\
 \det C_k&=\det(I_n+R^{2M})>0,\\
 f(z)&=\det C_k\det(I_r+F\diag(z)),\qquad
 a_J=\det C_k\det F[J,J].
 \end{split}
 \label{d:cyclic-source-identities}
 \end{equation}
 All these matrices are Gaussian rational when the input and $\delta$
 are rational. Here $F[J,J]$ denotes the principal submatrix with row
 and column set $J$, and its empty determinant is $1$.
\end{lemma}

Appendix~\ref{a:cyclic-proof} proves the determinant identity and the
partition independence. In the construction for this section we use
one chunk per macrostep, so $k=M$ and $C=C_M$ has dimension $nM$.
The entries of $F$ also have the useful explicit form
\begin{equation}
 \begin{split}
 Q_R&=R^{2M},\qquad P_R=Q_R(I_n+Q_R)^{-1},
       \qquad t_j=2(M-1-m(j)),\\
 F_{j\ell}&=v_j^\dagger R^{t_j}
       (P_R-\mathbf1_{\{j<\ell\}}I_n)R^{-t_\ell}u_\ell.
 \end{split}
 \label{d:chronological-kernel}
\end{equation}
In particular, events at the same macro time still have an order.
The term $-I_n$ in \eqref{d:chronological-kernel} records this order
and is present even when their endpoints are nonorthogonal.

\begin{proposition}[Nonnegative determinantal representation]
 \label{d:positive-sum}
 With $\gamma^J=\prod_{j\in J}\gamma_j$,
 \begin{equation}
 Z_{\mathrm{disc}}
   =|\det C|^2\mathcal W,\qquad
 \mathcal W=\sum_{J\subseteq[r]}
          \gamma^J|\det F[J,J]|^2\geq1.
 \label{d:positive-determinant-sum}
 \end{equation}
\end{proposition}
\begin{proof}
 Expand the event factors in $\Tr(G^2PP^\dagger)^M$.
 Choosing $J$ inserts exactly the same event set in the two components.
 In the first component its trace is $a_J$; all matrices and endpoints
 in the second component are their complex conjugates, so its trace is
 $\bar a_J$. The scalar interaction weight is $\gamma^J$.
 Consequently $Z_{\mathrm{disc}}=\sum_J\gamma^J|a_J|^2$.
 Substituting Lemma~\ref{d:cyclic-source} proves the formula. The
 empty event set contributes $1$ to $\mathcal W$.
\end{proof}

If $s=0$, there are no interaction events and
$Z_{\mathrm{disc}}=|\det(I_n+R^{2M})|^2$ is computed exactly.
For $s\geq1$, Proposition~\ref{d:positive-sum} gives a finite
nonnegative sum but does not yet give an efficient way to evaluate it.
The next section supplies an approximate counting algorithm.


\section{Polynomial-time approximation by log-concave counting}
\label{sec:counting}

The reduction in Section~\ref{sec:positive} leaves a nonnegative
determinant sum to be counted. We show that it is a direct instance
of the Chen--Liu theorem, proving the polynomial-time assertion in
Theorem~\ref{p:main}. Section~\ref{c:complementary-section}
encodes the normalized sum $\mathcal W$ as a complementary coefficient
of two polynomials and proves their log-concavity.
Section~\ref{c:oracles-section} verifies the exact coefficient oracles,
support seeds, and coefficient-range bounds, and applies the counting
theorem to approximate the discrete partition function in polynomial
bit time. Section~\ref{c:energy-section} combines this relative
approximation with the transfer and thermal estimates to prove the
ground-energy guarantee.

\subsection{Complementary columns and log-concavity}
\label{c:complementary-section}

Put $\mathcal A=[I_r\mid F]$. Label the identity columns by
$c_1,\ldots,c_r$ and the columns of $F$ by $b_1,\ldots,b_r$, and
write
\[
 \mathcal U=\{c_j,b_j:j\in[r]\}.
\]
These are column labels, not fermionic operators. For $S,T\subseteq
\mathcal U$, define
\begin{equation}
 \begin{split}
 \mu(S)&=
   \begin{cases}|\det\mathcal A_{:,S}|^2,&|S|=r,\\
                 0,&\text{otherwise},\end{cases}\\
 \nu(T)&=
   \begin{cases}\displaystyle\prod_{j:\,c_j\in T}\gamma_j,
       &|T\cap\{c_j,b_j\}|=1\text{ for every }j,\\
       0,&\text{otherwise}.
   \end{cases}
 \end{split}
 \label{c:weights}
\end{equation}
For any nonnegative subset weight $\omega$, its generating polynomial
is $g_\omega(x)=\sum_S\omega(S)x^S$, where $x^S=\prod_{u\in S}x_u$.
A nonzero polynomial with nonnegative coefficients is called
log-concave if its logarithm is concave on the positive orthant.

\begin{proposition}[Log-concave complementary representation]
 \label{c:representation}
 The polynomials
 \begin{equation}
 g_\mu(x)=\det(\mathcal A\diag(x)\mathcal A^\dagger),\qquad
 g_\nu(x)=\prod_{j=1}^r(\gamma_jx_{c_j}+x_{b_j})
 \label{c:generating-polynomials}
 \end{equation}
 are homogeneous of degree $r$, multiaffine, and log-concave, with
 nonnegative rational coefficients. Here $x$ has $2r$ coordinates,
 one per column label, so $\diag(x)$ is a $2r\times2r$ diagonal
 matrix. Moreover,
 \begin{equation}
 \mathcal W=[x^{\mathcal U}]g_\mu(x)g_\nu(x)
       =\sum_{S\subseteq\mathcal U}\mu(S)\nu(\mathcal U\setminus S).
 \label{c:complementary-coefficient}
 \end{equation}
\end{proposition}
\begin{proof}
 Cauchy--Binet expands the first determinant as
 $\sum_{|S|=r}|\det\mathcal A_{:,S}|^2x^S$. Expanding the second
 expression selects one label from each pair. This proves the
 generating-polynomial formulas, nonnegativity, homogeneity, and
 multiaffinity.

 A nonzero term of the complementary convolution has exactly one
 selected label from each pair in $S$. Let $J=\{j:b_j\in S\}$.
 Expanding along the selected identity columns gives
 $|\det\mathcal A_{:,S}|^2=|\det F[J,J]|^2$, while
 $\nu(\mathcal U\setminus S)=\gamma^J$. The correspondence between
 such $S$ and $J$ is bijective, proving
 \eqref{c:complementary-coefficient}.

 Since $\mathcal A$ contains $I_r$, the matrix
 $Q(x)=\mathcal A\diag(x)\mathcal A^\dagger$ is positive definite
 for $x>0$. For a real direction $h$, put
 $D_h=\mathcal A\diag(h)\mathcal A^\dagger$. Then
 \[
  \left.\frac{d^2}{dt^2}\log\det(Q(x)+tD_h)\right|_{t=0}
  =-\Tr\!\left[(Q(x)^{-1/2}D_hQ(x)^{-1/2})^2\right]\leq0.
 \]
 The matrix inside the square is Hermitian. Thus $g_\mu$ is
 log-concave. Finally,
 $\log g_\nu=\sum_j\log(\gamma_jx_{c_j}+x_{b_j})$ is a sum of
 concave functions.
\end{proof}

The two auxiliary polynomials are the objects whose log-concavity
is needed. No Hermiticity of $F$ or log-concavity of the original
single-component source polynomial $f$ is assumed.

\subsection{Counting oracles and the bit model}
\label{c:oracles-section}

For a nonzero weight $\omega$, define its positive coefficient range
by
\[
 \mathcal R_\omega=
 \frac{\max\{\omega(S):\omega(S)>0\}}
      {\min\{\omega(S):\omega(S)>0\}}.
\]
Zero coefficients are excluded. We use the following form of the
homogeneous counting theorem of Chen and Liu
\cite[Theorem~5.1 and Appendix~A.5, Remark~A.4]{ChenLiu2026}.

\begin{theorem}[Chen--Liu log-concave counting]
 \label{c:chen-liu}
 Let $\omega_1,\omega_2$ be nonzero nonnegative rational weights on
 subsets of a $2r$-element set $\mathcal V$. Suppose their generating
 polynomials are homogeneous of degree $r$, multiaffine, and
 log-concave. Given exact coefficient oracles, one support set for
 each weight, and upper bounds on their positive coefficient ranges,
 there is a randomized algorithm which, for rational $0<\eta\leq1$,
 approximates
 \[
 Z=\sum_S\omega_1(S)\omega_2(\mathcal V\setminus S)
 \]
 to relative error $\eta$ with probability at least $3/4$.
 Its arithmetic work and number of oracle calls are polynomial in
 $r,\eta^{-1}$ and the logarithms of the supplied coefficient-range
 bounds. If the oracles have polynomial bit complexity and return
 rationals of polynomial encoding length, the algorithm has a
 polynomial-time bit implementation on every execution, with the
 input encoding lengths included among the parameters.
\end{theorem}

\begin{proposition}[Relative approximation of the discrete partition function]
 \label{c:relative-counting}
 For rational transfer data satisfying $\delta\lambda\leq1/64$ and
 rational $0<\eta\leq1$, a classical randomized algorithm returns
 a nonnegative rational $\widehat Z_{\mathrm{disc}}$ such that
 \begin{equation}
 \Pr\bigl[(1-\eta)Z_{\mathrm{disc}}
       \leq\widehat Z_{\mathrm{disc}}
       \leq(1+\eta)Z_{\mathrm{disc}}\bigr]\geq3/4.
 \label{c:relative-guarantee}
 \end{equation}
 Its bit running time on every execution is polynomial in
 $n,s,M,\eta^{-1}$ and the encoding lengths of the input and $\delta$.
\end{proposition}
\begin{proof}
 The case $s=0$ is an exact determinant. Otherwise construct
 $C,\widehat U,\widehat V,L,F$ using Lemma~\ref{d:cyclic-source}.
 Their dimensions are polynomial in $n,s,M$. Entries of the initial
 matrices have polynomial encoding length. Clearing denominators,
 the determinant and adjugate formulas bound the entries of $C^{-1}$
 and $F$ by ratios of determinants of polynomial-size integer
 matrices. Hadamard's inequality therefore gives polynomial bounds
 on their encoding lengths. Exact rational elimination computes
 them in polynomial bit time.

 For $\mu$, the coefficient oracle computes a determinant and its
 squared modulus. For $\nu$, it checks that exactly one label is
 selected from each pair and then multiplies the indicated $\gamma_j$.
 These are exact rational procedures, including exact tests for zero
 minors. Support seeds are
 \[
  S_c=\{c_j:j\in[r]\},\qquad S_b=\{b_j:j\in[r]\},\qquad
  \mu(S_c)=\nu(S_b)=1.
 \]

 To make the coefficient-range bound explicit, suppose the real and
 imaginary numerators and denominators in $\mathcal A$ have absolute
 value at most $2^{b_F}$, with positive denominators. Their product
 gives a common positive denominator $Q$ with
 $\log_2Q\leq4r^2b_F$. A nonzero determinant of $r$ columns has a
 nonzero Gaussian-integer numerator after multiplication by $Q^r$.
 Hadamard's inequality then gives
 \[
  \mu(S)>0\ \Longrightarrow\quad
  Q^{-2r}\leq\mu(S)\leq(2r)^r2^{2rb_F},\qquad
  \mathcal R_\mu\leq Q^{2r}(2r)^r2^{2rb_F}.
 \]
 If positive $\gamma_j$ have numerators and denominators at most
 $2^{b_\gamma}$, then $\mathcal R_\nu\leq2^{2rb_\gamma}$.
 The logarithms of both bounds are polynomial in the input size
 and the transfer dimensions. Proposition~\ref{c:representation}
 verifies the remaining hypotheses of Theorem~\ref{c:chen-liu}.
 Apply that theorem to $\mathcal W$, replace a negative reported
 value by zero, and multiply by the exact positive rational
 $|\det C|^2$. Clipping changes no successful relative approximation.
\end{proof}

Arbitrarily small positive couplings enter these estimates through
their encoding lengths. There is no additional numerical lower-bound
promise on a nonzero coupling or a nonzero minor. This first algorithm
uses the full Chen--Liu counting theorem; Section~\ref{sec:fast}
will give an explicit counting procedure tailored to the present
determinantal representation.

\subsection{From relative counting to the ground energy}
\label{c:energy-section}

Use the parameters in \eqref{c:parameters}. Approximate $Z_{\mathrm{disc}}$ with relative error $1/4$ using
Proposition~\ref{c:relative-counting}. If the reported value is
nonpositive, return $0$. Otherwise compute a rational $\ell$ with
\[
 |\ell-\log\widehat Z_{\mathrm{disc}}|
 \leq\beta\varepsilon/16
\]
and return $\widehat E=c-\ell/\beta$. The scalar calculation has
polynomial bit complexity by Lemma~\ref{a:scalar-logarithm}.

\begin{proof}[Proof of the polynomial-time assertion in Theorem~\ref{p:main}]
 With probability at least $3/4$, the counting call satisfies
 $\frac34Z_{\mathrm{disc}}\leq\widehat Z_{\mathrm{disc}}
 \leq\frac54Z_{\mathrm{disc}}$. On this event its output is positive
 and
 $|\log\widehat Z_{\mathrm{disc}}-\log Z_{\mathrm{disc}}|
 \leq\log(4/3)$.
 Proposition~\ref{d:transfer-error} and the variational principle
 imply
 $|E_0(H_0)-E_0(H_{\mathrm{eff}})|\leq10\delta\lambda^2$.
 Applying Lemma~\ref{p:thermal} to $H_{\mathrm{eff}}$ on the full
 $4^n$-dimensional space gives
 \begin{equation}
 \begin{split}
 |\widehat E-E_0(H)|
 &\leq10\delta\lambda^2+\frac{n\log4}{\beta}
       +\frac{\log(4/3)}{\beta}+\frac\varepsilon{16}\\
 &\leq\frac\varepsilon{20}
       +\frac{\varepsilon\log4}{16}
       +\frac{\varepsilon\log(4/3)}{16n}
       +\frac\varepsilon{16}<\varepsilon.
 \end{split}
 \label{c:energy-budget}
 \end{equation}
 The four terms account for the transfer approximation, thermal
 approximation, relative counting error, and scalar logarithm error.

 Finally,
 $M=O(n\lambda^2\varepsilon^{-2})$ and
 $r=O(sn\lambda^2\varepsilon^{-2})$.
 Proposition~\ref{c:relative-counting} and
 Lemma~\ref{a:scalar-logarithm} give a bit bound polynomial in
 $n,s,\lambda,\varepsilon^{-1}$ and the input encoding length on
 every execution, including counting failures. The scalar shift
 $c$ is added exactly; its magnitude contributes only through its
 encoding length. The relative counting call is the sole randomized
 step, so the success probability remains at least $3/4$.
\end{proof}

This argument estimates the minimum over the full Fock space. The
algorithm chooses its own inverse temperature $\beta$ and uses no
assumption on a physical spectral gap. Its dependence on the numerical
strength $\lambda$ is explicit; a binary-encoded exponentially large
strength is not treated as a polynomial-size numerical parameter.

\section{A faster counting algorithm}
\label{sec:fast}

Section~\ref{sec:counting} proves the polynomial-time guarantee by
applying the general Chen--Liu counting theorem. We now give a second
proof through a specialized counter for the same determinant sum and
the same full Hamiltonian class. A finer analysis of its sampling and
matrix operations yields the explicit arithmetic bound. The algorithm
uses the exchange
Poincar\'e inequality of Chen and Liu, together with a cooling schedule
and matrix implementation adapted to the present model. Two properties
are essential: conjugate pairing turns a deformation of the determinant
sum into a positive transfer trace, and rank-one insertions admit a
sparse representation whose blocks have size $O(n)$.

We retain the transfer parameters of Section~\ref{sec:positive}:
$\beta=16n/\varepsilon$, $M=\lceil200\beta\lambda^2/\varepsilon\rceil$,
$\delta=\beta/M$, and $R=I_n-\delta q/2$. There are $r=2sM$
insertion events. At an adjoint event the endpoints of its channel are
interchanged. Events are ordered first by macrostep and then by their
position in the ordered product. The matrix $F$ and positive free
determinant $\det C=\det(I_n+R^{2M})$ are those of
Lemma~\ref{d:cyclic-source}. In particular,
\begin{equation}
 Z_{\mathrm{disc}}=|\det C|^2\mathcal W,\qquad
 \mathcal W=\sum_{J\subseteq[r]}|\det F[J,J]|^2\gamma^J.
 \label{f:target}
\end{equation}
Here $\gamma_j=\delta g_\alpha/2$ if event $j$ belongs to channel
$\alpha$. If all channels vanish, $\mathcal W=1$ and no randomized
counting is necessary. We henceforth discard zero channels and assume
$s\geq1$.

\subsection{Rational normalization and counting weights}
\label{f:normalization}

We first rescale the channel endpoints to bound their norms while preserving
every transfer factor. We then round the counting weights to rational squares
so that Section~\ref{f:pair-source} can weight the two endpoints separately. For each channel choose
$d_\alpha,e_\alpha\in\{2^k:k\in\mathbb Z\}$ with
\[
 \norm{u_\alpha}_2\leq d_\alpha<2\norm{u_\alpha}_2,
 \qquad \norm{v_\alpha}_2\leq e_\alpha<2\norm{v_\alpha}_2,
\]
and replace
\[
 u_\alpha\leftarrow u_\alpha/d_\alpha,\quad
 v_\alpha\leftarrow v_\alpha/e_\alpha,\quad
 g_\alpha\leftarrow g_\alpha d_\alpha^2e_\alpha^2.
\]
The Hamiltonian and $\lambda$ are unchanged. Comparisons of squared
norms with powers of four determine the exponents using polynomially
many bit operations. Under this rescaling, \(B_\alpha\) is divided by \(d_\alpha^2e_\alpha^2\), so \(g_\alpha B_\alpha\) and the transfer factor \(I+(\delta g_\alpha/2)B_\alpha\) are unchanged. The channel contribution \(g_\alpha\|u_\alpha\|_2^2\|v_\alpha\|_2^2\) to \(\lambda\) is also unchanged. Write
$p_\alpha=\norm{u_\alpha}_2^2$ and
$q_\alpha=\norm{v_\alpha}_2^2$ for the squared norms of the rescaled endpoints.
These scalars, distinct from the matrix $q$, satisfy
\begin{equation}
 \tfrac14<p_\alpha,q_\alpha\leq1,\qquad
 p_\alpha+q_\alpha\leq\tfrac52\sqrt{p_\alpha q_\alpha},\qquad
 \sum_\alpha g_\alpha\leq16\lambda.
 \label{f:norm-bounds}
\end{equation}
Consequently, using $\delta\lambda\leq1/200$ from \eqref{c:parameters},
$\gamma_\alpha=\delta g_\alpha/2\leq8\delta\lambda\leq1/25<1/4$.

Let $0<\eta\leq1$ be the required relative counting accuracy and put
$N=2r$. We replace every positive $\gamma_\alpha$ by a rational square
$\widetilde\gamma_\alpha=s_\alpha^2$ satisfying
\begin{equation}
 0<s_\alpha\leq\sqrt{\gamma_\alpha},\qquad
 1-s_\alpha/\sqrt{\gamma_\alpha}\leq\eta/(10^5N).
 \label{f:square-accuracy}
\end{equation}
The same $s_\alpha$ is used at every repetition and adjoint of the
channel. An explicit binary construction is given in
Appendix~\ref{fa:finite-bits}. Every monomial in \eqref{f:target}
contains at most $r$ factors, so its rounded weight, and hence the
whole sum, changes by a factor between $e^{-\eta/1000}$ and
$e^{\eta/1000}$. Denote the rounded target by $W_0$. A relative
$\eta/2$ approximation of $W_0$ therefore suffices for relative
$\eta$ approximation of $\mathcal W$.

Only auxiliary counting weights are rounded: $q$, $R$, $H$, and the
unrounded transfer error estimate remain fixed. Integer division,
integer square root, dyadic exponents, and integer control parameters
are charged to a separate polynomial preprocessing bit cost
$P_{\mathrm{pre}}$. The arithmetic bounds below count
$+,-,\times,\div$ on rational numbers after this preprocessing, with
rational comparisons and index bookkeeping of the same order.

For $x\geq1$, write
$\ell_2(x)=\min\{k\in\mathbb Z_{\geq0}:2^k\geq x\}$;
integer comparisons determine this value from a binary rational input.
Define the exactly rational parameters
\begin{equation}
 h_\alpha=\widetilde\gamma_\alpha p_\alpha q_\alpha,\qquad
 B_{\mathrm{src}}=2M\sum_\alpha h_\alpha\leq\beta\lambda,
 \qquad a=\max\{1,\lceil\sqrt{rB_{\mathrm{src}}}\rceil\}.
 \label{f:effective-size}
\end{equation}
For analysis only, let $Q_{\mathrm{src}}=\sum_\alpha\sqrt{h_\alpha}$
and $a_0=MQ_{\mathrm{src}}$. Cauchy--Schwarz gives
$2a_0\leq\sqrt{rB_{\mathrm{src}}}\leq a$.
Also $1\leq a\leq r+1\leq N$, since
$\sum_\alpha h_\alpha\leq\delta\lambda/2\leq1/400$ and
$B_{\mathrm{src}}\leq r/400$.

\subsection{A pair source for noncommuting channels}
\label{f:pair-source}

For an event set $I$, put $s_I=\prod_{j\in I}s_j$, where $s_j$ is
its channel's rounded square root. Introduce the polynomial
\begin{equation}
 W_\theta=\sum_{\substack{I,J\subseteq[r]\\|I|=|J|}}
    |\det F[I,J]|^2s_Is_J\theta^{|I\triangle J|},\qquad\theta\geq0.
 \label{f:source-polynomial}
\end{equation}
Only even powers of $\theta$ occur, since $|I|=|J|$, and all coefficients
are nonnegative. At $\theta=0$ only terms with $I=J$ survive, so
$W_\theta\geq W_0\geq1$.
To control its growth, use auxiliary tensor operators
\[
 \mathsf C_u=a^\dagger(u)\otimes\overline{a^\dagger(u)},\qquad
 \mathsf A_v=a(v)\otimes\overline{a(v)},
\]
and define
\[
 L_\alpha(\theta)=I+\widetilde\gamma_\alpha B_\alpha
       +\theta s_\alpha(\mathsf C_{u_\alpha}+\mathsf A_{v_\alpha}),
 \qquad X_\theta=G\prod_{\alpha=1}^sL_\alpha(\theta),
 \quad G=\Gamma(R)\otimes\Gamma(\overline R).
\]
These are operators on the auxiliary tensor product
$\F_n\otimes\F_n$. The displayed creation term is used as a tensor
operator; identifying it with creation of a physical pair would require
the parity convention discussed in Section~\ref{sec:prelim}.

\begin{lemma}[Pair-source identity]
\label{f:source-identity}
For every $\theta\geq0$,
\begin{equation}
 W_\theta=\frac{\Tr[(X_\theta X_\theta^\dagger)^M]}{|\det C|^2}.
 \label{f:source-trace}
\end{equation}
\end{lemma}
\begin{proof}
At an event choose both endpoints, only its creation endpoint, only
its annihilation endpoint, or neither. A single-component trace vanishes
unless the numbers of chosen creation and annihilation endpoints agree.
When their index sets are $I,J$, respectively, the finite CAR recursion
gives the trace $\epsilon_{I,J}\det C\det F[J,I]$, with an ordering
sign $\epsilon_{I,J}\in\{1,-1\}$. The chronological kernel includes
the anticommutator of nonorthogonal endpoints at the same macrostep;
its explicit derivation is in Appendix~\ref{fa:wick}. The other component
has exactly the conjugate ordered word. Its trace removes the sign
and gives the squared modulus. Both endpoints contribute $s_j^2$,
and an unmatched endpoint contributes $\theta s_j$. After interchanging
the names $I,J$, their sum is \eqref{f:source-polynomial}. The adjoint
half reverses channel order and swaps endpoints, exactly as required
by $X_\theta X_\theta^\dagger$.
\end{proof}

The factors $L_\alpha(\theta)$ need not commute or be Hermitian.
Nevertheless, their positive product with its adjoint suffices for a
uniform growth estimate.

\begin{lemma}[Growth of the pair source]
\label{f:source-growth}
For all $\theta\geq0$,
\begin{equation}
 0\leq\frac{d}{d\theta}\log W_\theta\leq8a.
 \label{f:growth-bound}
\end{equation}
\end{lemma}
\begin{proof}
Put $H_{\mathrm{src}}=\sum_\alpha h_\alpha\leq1/400$.
The norm identities for linear fermionic operators and
\eqref{f:norm-bounds} give
\[
 \norm{L_\alpha(\theta)-I}
 \leq h_\alpha+\tfrac52\theta\sqrt{h_\alpha},\qquad
 \norm{L_\alpha'(\theta)}\leq\tfrac52\sqrt{h_\alpha}.
\]
First suppose $\theta Q_{\mathrm{src}}\leq1/64$. The sum of the
increment norms is less than $1/8$. A product of any suffix of the
factors therefore has condition number less than two: if the local
increment norms are $e_\alpha$, its condition number is at most
$\prod_\alpha(1+e_\alpha)/(1-e_\alpha)<2$.
Each inverse has norm less than $8/7$. The product rule, written as
an ordered right logarithmic derivative, consequently yields
\[
 \norm{X_\theta^{-1}X_\theta'}
 =\norm{\left(\prod_\alpha L_\alpha\right)^{-1}
                  \left(\prod_\alpha L_\alpha\right)'}
 \leq6Q_{\mathrm{src}}.
\]
Indeed every summand is a conjugate of
$L_\alpha^{-1}L_\alpha'$ by the factors to its right.
For $S_\theta=X_\theta X_\theta^\dagger$, put $K=X_\theta^{-1}X_\theta'$.
Then $S_\theta'=X_\theta(K+K^\dagger)X_\theta^\dagger$.
Since $\norm{K+K^\dagger}\leq2\norm{K}$, this implies
$\pm S_\theta'\preceq12Q_{\mathrm{src}}S_\theta$.
Taking the positive trace pairing with $S_\theta^{M-1}$ in
$\frac{d}{d\theta}\Tr S_\theta^M=M\Tr(S_\theta^{M-1}S_\theta')$
proves $| (\log W_\theta)'|\leq12a_0$ in this regime.

For arbitrary $\theta$, no inverse is needed. The free-generator bound
\eqref{a:free-logarithm} gives
$\norm{G}\leq\exp(3\delta\lambda/5)$, so the product norm satisfies
\[
 \norm{X_\theta}\leq
 \exp\bigl(3\delta\lambda/5+H_{\mathrm{src}}
                       +(5/2)\theta Q_{\mathrm{src}}\bigr).
\]
The dimension is $4^n$ and $|\det C|^2\geq1$. Hence
\begin{equation}
 0\leq\log W_\theta
 \leq n\log4+(11/5)\beta\lambda+5a_0\theta
 \leq3\beta\lambda+5a_0\theta.
 \label{f:global-trace-bound}
\end{equation}
Here $\beta\lambda\geq16n$. Because the coefficients of $W$ are
nonnegative, $\Phi(x)=\log W_{e^x}$ is convex. Set
$m(\theta)=\frac12\theta(\log W_\theta)'$. The forward secant gives
\[
 2m(\theta)\leq
 \frac{\log W_{2\theta}-\log W_\theta}{\log2}
 \leq\frac{3\beta\lambda+10a_0\theta}{\log2}.
\]
In the remaining regime $\theta Q_{\mathrm{src}}>1/64$,
$a_0\theta>M/64$, whereas
$3\beta\lambda\leq3M/200<M/64$.
Thus $m(\theta)<8a_0\theta$, since $11/\log 2<16$, and
$(\log W_\theta)'\leq16a_0\leq8a$.
In the first regime the bound $12a_0\leq6a$ suffices.
Nonnegativity of the coefficients gives monotonicity, and continuity
gives the bound at $\theta=0$.
\end{proof}

\subsection{A collapsed exchange chain}
\label{f:collapse}

We now construct a Markov chain for the deformed minor sum $W_\theta$.
We first express its weights as a marginal of the two-set law used by Chen and Liu.
This representation lets us update a single set while retaining a Poincar\'e
bound from the joint exchange chain.

Set $\mathcal A=[I_r\mid F]$ and label its columns $c_j,b_j$.
Use positive rational fields $\zeta_{c_j}=1/s_j$ and
$\zeta_{b_j}=s_j$. On subsets of the $N=2r$ labels define
\[
 \mu(S)=|\det\mathcal A_{:,S}|^2\quad(|S|=r),\qquad
 \nu(T)=\prod_{j:c_j\in T}\widetilde\gamma_j
 \quad\bigl(|T\cap\{c_j,b_j\}|=1\text{ for every }j\bigr),
\]
and zero otherwise. Their generating polynomials are
\[
 g_\mu(z)=\det(\mathcal A\diag(z)\mathcal A^\dagger),\qquad
 g_\nu(z)=\prod_j(\widetilde\gamma_jz_{c_j}+z_{b_j}).
\]
Both are homogeneous, multiaffine, and log-concave on the positive
orthant, by the argument of Section~\ref{sec:counting}.
For $t>0$ consider the joint law
\begin{equation}
 \pi_t(S,T)=\mathcal Z_t^{-1}
      \mu(S)\nu(T)\zeta^S\zeta^Tt^{|S\cap T|}.
 \label{f:joint-law}
\end{equation}
We use the following consequence of the exchange inequality
\cite[Lemma~5.5 and Eq.~(19)]{ChenLiu2026}. For $0<t\leq1$, the
reversible generator with first-component off-diagonal rates
\begin{equation}
 q((S,T),(S-x+y,T))=
 t^{-\mathbf1_{\{x\in T\}}}
 \frac{\zeta_y\mu(S-x+y)}
      {\sum_{w\notin S-x}\zeta_w\mu(S-x+w)},
 \quad x\in S,\ y\notin S,
 \label{f:joint-rates}
\end{equation}
and the analogous second-component rates satisfies
$\operatorname{Var}_{\pi_t}f\leq\mathcal E_{Q_t}(f,f)$, where
\[
 \mathcal E_{Q_t}(f,f)=\tfrac12\sum_{x,y}
                  \pi_t(x)q(x,y)(f(y)-f(x))^2.
\]
The reinsertion denominator includes $w=x$, so is positive on the
support. This is the sole external mixing theorem in this section.

For a supported basis $S$ let $D(S)$ be the number of pairs
$\{c_j,b_j\}$ containing two selected labels. There are also $D(S)$
empty pairs. Define
\begin{equation}
 \rho_\theta(S)\ \propto\ |\det\mathcal A_{:,S}|^2\zeta^S
                                  \theta^{2D(S)}.
 \label{f:marginal-law}
\end{equation}
Writing $I=\{j:c_j\notin S\}$ and $J=\{j:b_j\in S\}$ gives
$|I|=|J|$, $2D(S)=|I\triangle J|$, and
$|\det\mathcal A_{:,S}|^2=|\det F[I,J]|^2$.
The normalizer in \eqref{f:marginal-law} is therefore
$W_\theta/\prod_js_j$.
Direct summation of $T$ in \eqref{f:joint-law} gives
\begin{equation}
 \mathcal Z_{u^2}=(1+u^2)^rW_{\theta(u)},\qquad
 \theta(u)=\frac{2u}{1+u^2}.
 \label{f:theta-map}
\end{equation}
To see this, a single pair contributes $1+t$ or
$\widetilde\gamma_j(1+t)$ after its fields are included; an empty
pair contributes $2s_j$, and a double pair contributes $2ts_j$.
Equal empty and double counts give the displayed factor.

\begin{lemma}[Moment and weighted-curvature bounds]
\label{f:moments}
For $0<\theta\leq2$,
\[
 m(\theta):=\mathbb E_{\rho_\theta}D\leq4a\theta,\qquad
 \mathbb E_{\rho_\theta}D^2\leq2^{14}(1+a\theta)^2.
\]
For every fixed $c_*>0$ and $\theta_-\geq c_*/a$,
\begin{equation}
 \int_{\theta_-}^{2}
       (N+a/\theta+\theta^{-2})\operatorname{Var}_{\rho_\theta}(D)
             \,d\log\theta
 \leq C_{c_*}[Na+a^2\log(2a)].
 \label{f:weighted-curvature}
\end{equation}
\end{lemma}
\begin{proof}
Differentiating the finite sum gives
$2m(\theta)=\theta(\log W_\theta)'$ and
$dm/d\log\theta=2\operatorname{Var}(D)$.
The first claim follows from Lemma~\ref{f:source-growth}.
For $z=[64(1+a\theta)]^{-1}$, that lemma also gives
\[
 \log\mathbb E e^{zD}
 =\log W_{\theta e^{z/2}}-\log W_\theta
 \leq8a\theta(e^{z/2}-1)<1/2.
\]
The inequality $y^2\leq2e^y$ proves the stated second moment.
For $w(\theta)=N+a/\theta+\theta^{-2}$, integration by parts gives
\[
 \begin{split}
 \int_{\theta_-}^{2}w\operatorname{Var}(D)d\log\theta
 &=\tfrac12\int_{\theta_-}^{2}w\,dm\\
 &\leq\tfrac12w(2)m(2)
    +\tfrac12\int_{\theta_-}^{2}
              m(\theta)(a\theta^{-2}+2\theta^{-3})d\theta\\
 &\leq C[Na+a^2\log(2/\theta_-)+a/\theta_-].
 \end{split}
\]
We dropped the nonpositive lower boundary term. The assumption on
$\theta_-$ now implies \eqref{f:weighted-curvature}.
\end{proof}

\begin{lemma}[Collapsed dynamics]
\label{f:collapsed-chain}
For $0<u\leq1$, averaging the first-component rates \eqref{f:joint-rates} over
$T$ conditional on $S$ gives a reversible generator $\overline Q_u$
for $\rho_{\theta(u)}$ with Poincar\'e constant at most one.
The chain
\begin{equation}
 P_u=I+(u^2/N)\overline Q_u
 \quad\text{has}\quad\operatorname{gap}(P_u)\geq g_u:=u^2/N.
 \label{f:gap}
\end{equation}
At each discrete-time step, choose one of the $r$ selected labels uniformly.
If it belongs to a singly occupied pair, activate with probability $u^2/(1+u^2)$;
if it belongs to a doubly occupied pair, activate with probability $(1+u^2)/4$.
Upon activation, delete the label and reinsert according to the
unpenalized weighted determinant conditional law.
The activation probability at $S$ is
\begin{equation}
 \frac{u^2}{1+u^2}
 +\frac{D(S)}r\frac{(1-u^2)^2}{2(1+u^2)}
 \leq u^2+D(S)/N.
 \label{f:activation-probability}
\end{equation}
\end{lemma}
\begin{proof}
The conditional probability that a selected label lies in $T$ is
$t/(1+t)$ for a single pair and $1/2$ for a double pair. Averaging
$t^{-\mathbf1_{\{x\in T\}}}$ gives removal multipliers
$2/(1+t)$ and $(1+t)/(2t)$, respectively. Reinsertion is independent
of $T$. For a function of $S$ alone, the joint variance and Dirichlet
form equal those of the collapsed law and generator. The imported
inequality therefore proves the collapsed Poincar\'e bound.
Scaling by $t/N$ gives the stated clock construction. Each displayed
activation probability is at most $1/2$, so $P_u$ is lazy and its
centered $L^2$ contraction is at most $1-g_u$.
Counting its single and double labels gives
\eqref{f:activation-probability}.
\end{proof}

\subsection{Rational cooling and importance sampling}
\label{f:cooling}

Section~\ref{f:collapse} constructs a Markov chain whose stationary law assigns
probabilities proportional to the minor weights at each fixed $0<\theta\leq1$,
and gives a lower bound on its spectral gap. Our goal, however, is to estimate
the normalizing constant $W_0$. Since $W_1$ can be computed exactly, we use
annealed importance sampling to estimate $W_0/W_1$: each path accumulates
weights as $\theta$ decreases from one, and a final restriction to $D=0$
selects the principal minors that contribute to $W_0$.

Fix $\kappa_0=2^{30}$ and let
\[
 u_*=\frac1{\kappa_0 a},\quad
 C_{\mathrm{cool}}=\kappa_0^4[Na+a^2(1+\ell_2(2a))],\quad
 w(u)=N+a/u+u^{-2}.
\]
Start at $u_0=1$. If $u_j>u_*$, choose a dyadic
$\alpha_j=2^{-\ell_j}$ satisfying
\[
 \frac{w(u_j)}{2C_{\mathrm{cool}}}<\alpha_j\le\frac{w(u_j)}{C_{\mathrm{cool}}},
 \qquad
 u_{j+1}=\max\{u_*,u_j(1-\alpha_j)\}.
\]
Stop at $u_{J_0}=u_*$. All quantities are rational or integer and exactly
computable; the integer and dyadic choices use the preprocessing
convention in Section~\ref{f:normalization}. The mesh has
\begin{equation}
J_0=\widetilde O(C_{\mathrm{cool}}/N)=\widetilde O(a),\qquad
 \sum_{j=1}^{J_0}w(u_j)=\widetilde O(C_{\mathrm{cool}}).
\label{f:mesh-sums}
\end{equation}
These follow by comparison with integrals in $d\log u$; the mesh
logarithmic step is comparable to $w(u)/C_{\mathrm{cool}}$, and $w$ changes by a
constant factor in a step. Here $a\le r+1\le N$, and the large mesh
constant ensures every step is small.

\begin{lemma}[Bridge bounds]
\label{f:bridge-lemma}
Put $\theta_j=\theta(u_j)$ and
$f_j(S)=(\theta_{j+1}/\theta_j)^{2D(S)}$ for $j<J_0$.
Let $a_j=\mathbb E_{\rho_{\theta_j}}f_j$. These normalizing ratios and $G_j$
are used only in the analysis; the algorithm evaluates the explicit
$f_j$. For the mesh above,
\begin{equation}
G_j:=f_j/a_j\le2,
 \qquad \sum_{j<J_0}\operatorname{Var}_{\rho_{\theta_j}}G_j\le1,
 \qquad 1+\chi^2(\rho_{\theta_j}\Vert\rho_{\theta_{j+1}})\le2.
\label{f:bridge-bounds}
\end{equation}
The terminal bridge is $f_{J_0}=\mathbf1_{\{D=0\}}$, whose mean is at
least $3/4$ and whose normalized version is bounded by $4/3$.
\end{lemma}
\begin{proof}
Write $\xi=\log\theta$, $\Phi(\xi)=\log W_{e^\xi}$, and
$\lambda_j=\xi_j-\xi_{j+1}$. Then $\Phi''(\xi)=4\operatorname{Var}(D)$ and
\begin{equation}
\log(1+\operatorname{Var} G_j)
 =\Phi(\xi_j)+\Phi(\xi_j-2\lambda_j)-2\Phi(\xi_j-\lambda_j)
 \le4\lambda_j\int_{\xi_j-2\lambda_j}^{\xi_j}\operatorname{Var}(D)d\xi.
\label{f:bridge-curvature}
\end{equation}
The derivative $d\log\theta(u)/d\log u=(1-u^2)/(1+u^2)$ belongs
to $[0,1]$ and increases during cooling. The logarithmic $u$-steps are
nondecreasing because the dyadic $\alpha_j$ are nondecreasing.
Thus the $\lambda_j$ are nondecreasing, except possibly for the
truncated final step. Each interval in \eqref{f:bridge-curvature} is contained in its own
mesh interval and the following one, with a possible terminal tail.
Only the last two forward intervals can extend below $\xi_{J_0}$, and both
remain above $\log(1/(2\kappa_0a))$ since $\lambda_j<\log 2$ and
$\theta_{J_0}\geq u_*$. Including these tails, the overlap multiplicity
is at most three. On each such interval,
\[
 \lambda_j\le 2w(u_j)/C_{\mathrm{cool}}\le Cw(e^\xi)/C_{\mathrm{cool}}.
\]
Use \eqref{f:weighted-curvature}, extending the lower limit to $1/(2\kappa_0 a)$, to bound the sum
in \eqref{f:bridge-curvature}. Keeping the dependence on the fixed endpoint constant
explicit, its weighted curvature integral is at most
$32\kappa_0[Na+a^2\log(2a)]$. Lemma~\ref{f:moments} gives
$m(\theta)\le4a\theta$, and the integration by parts above gives
this numerical bound directly. The mesh step obeys
$\lambda_j\le6/\kappa_0^2$; the comparison and overlap constants in \eqref{f:bridge-curvature}
are less than $2^{10}$. Division by $C_{\mathrm{cool}}$ therefore makes the sum
of logarithmic relative variances less than $2^{15}/\kappa_0^3$.
This also bounds the sum of relative variances by one.

For a reverse bridge, the logarithm of its second moment is a
curvature integral on $[\xi_{j+1},\xi_{j+1}+2\lambda_j]$. The same
weighted-integral bound applies to each individual interval. An
interval above $\xi=0$ lies below $\log2$, so the extension in \eqref{f:weighted-curvature}
also covers the first reverse comparison. This proves the last
inequality in \eqref{f:bridge-bounds}, without needing a sum of reverse second moments.

Jensen gives $a_j\ge\exp(-2\lambda_j m(\theta_j))$.
Since $\theta_j\asymp u_j$,
\[
 \lambda_jm(\theta_j)
 \le C a u_jw(u_j)/C_{\mathrm{cool}}
 \le C a[N+a+\kappa_0 a]/C_{\mathrm{cool}},
\]
which is less than $1/10$ for the specified $\kappa_0$. Hence $G_j\le2$.
Finally $\log(W_{\theta_{J_0}}/W_0)\le Ca\theta_{J_0}\le C'/\kappa_0$,
so the terminal success mass is at least $3/4$.
\end{proof}

Lemma~\ref{f:bridge-lemma} controls each bridge under its equilibrium law.
Along an actual importance path, successive states are correlated. The next
lemma shows that invariance preserves the mean exactly and that spectral
contraction controls the additional second-moment cost.

\begin{lemma}[An AIS second-moment statement with spectral mixing]
\label{f:ais-second-moment}
Let $\pi_0,\ldots,\pi_{J_0}$ be finite laws with common support and
normalized bridges $G_j=\pi_{j+1}/\pi_j\le R_{\mathrm{br}}$. Let $\mathsf P_j$ be reversible for
$\pi_j$ and contract centered $L^2(\pi_j)$ by at most $e$.
Start $X_0\sim\pi_0$, and after evaluating $G_j(X_j)$ move by
$\mathsf P_{j+1}$ to obtain $X_{j+1}$. If
$\sum_j\operatorname{Var}_{\pi_j}G_j\le V$, then, for sufficiently small $e$,
\[
 \mathbb E\prod_jG_j(X_j)=1,\qquad
 \mathbb E\left(\prod_jG_j(X_j)\right)^2
 \le\exp(V+C_{\mathrm{br}} J_0 e).
\]
A final bridge to a restricted support, bounded by the same $R_{\mathrm{br}}$,
may be evaluated without a subsequent transition. Its relative
variance is simply added to $V$.
\end{lemma}
\begin{proof}
The first assertion is the usual exact Feynman--Kac telescoping,
using only invariance, so requires no mixing approximation.
For completeness, let $h_j$ be the density relative to $\pi_j$ of
the squared, normalized weight accumulated before evaluating $G_j$.
Then $h_0=1$ and
$h_{j+1}=\mathsf P_{j+1}(h_jG_j)$.
Writing $m_j=\mathbb E_{\pi_j}h_j$ and
$z_j=\|h_j-m_j\|_{2,\pi_j}$ gives
\[
 m_{j+1}\le(1+v_j)m_j+R_{\mathrm{br}}^2z_j,
 \qquad z_{j+1}\le e R_{\mathrm{br}}^{3/2}(m_j+z_j),
 \quad v_j=\operatorname{Var}_{\pi_j}G_j.
\]
The second inequality uses
$\|h_jG_j\|_{2,\pi_{j+1}}^2
 =\mathbb E_{\pi_j}h_j^2G_j^3\le R_{\mathrm{br}}^3\|h_j\|_{2,\pi_j}^2$.
Take $2eR_{\mathrm{br}}^{3/2}\le1$. Induction over the stages with
transitions gives $m_j\le U_j$ and
$z_j\le2eR_{\mathrm{br}}^{3/2}U_j$, where
\[
 U_0=1,\qquad
 U_{j+1}=\left(1+v_j+
        2eR_{\mathrm{br}}^{7/2}\mathbf1_{\{j\ge1\}}\right)U_j.
\]
The first step needs no mixing-error charge: $z_0=0$ gives
$m_1=1+v_0$. A terminal restricted-support bridge uses only the first
recurrence, with its variance added to $V$; no density or contraction
on the restricted support is needed. Its possible extra charge has
index $j=J_0$. Thus at most $J_0$ charges remain, with indices
$1,\ldots,J_0$, including this terminal bridge when present. This
proves the stated exponent with
$C_{\mathrm{br}}=2R_{\mathrm{br}}^{7/2}$.
If $J_0=0$ and a terminal bridge is present, its second moment is
exactly $1+v_0\le \exp(v_0)$, so there is no mixing-error term.
\end{proof}

\subsubsection{The path estimator}
For each independent path draw one $S_0$ from the initial DPP. At
stage $j<J_0$, multiply the weight by $f_j(S_j)$ and then run the
collapsed chain at $u_{j+1}$ for $k_{j+1}$ clocks. At the endpoint
multiply by $\mathbf1_{\{D=0\}}$. Multiplying the average by $W_1$ estimates $W_0$; the complete algorithm below also includes $|\det C|^2$ and the rounding-error allowance. Use
$L=\lceil2^{14}\eta^{-2}\rceil$ independent path weights.
Lemma~\ref{f:ais-second-moment} and \eqref{f:bridge-bounds}
show unbiasedness for $W_0$ in the ideal uncapped chain and a universal relative second-moment
bound, provided each block contracts by at most $1/(C J_0)$.
The total $V$ is at most $4/3$, and the chosen mixing precision makes
the normalized second moment less than four. Chebyshev therefore
gives relative error $\eta/2$ with failure probability below $1/64$.
This is an average of independent paths;
there are no independent restarts at intermediate stages.

\subsubsection{Mixing budgets and activation counts}
Choose a logarithmic integer $p$ as follows. Set
\[
 C_{\rm mix}=1+J_0+\sum_{j=1}^{J_0 }g_{u_j}^{-1},\quad
 V_0=2^{30}(J_0+1)^2C_{\rm mix},\quad p=2\ell_2(V_0)+10,
 \quad k_j=\lceil p/g_{u_j}\rceil,
 \quad \mathcal T=\sum_jk_j.
\]
Then $\mathcal T\le pC_{\rm mix}$ and
$2^p\ge2^{20}(J_0+1)^2(\mathcal T+1)$. The block contraction is at most
$e^{-p}\le2^{-p}$, satisfying the AIS lemma.
The integer mixing budgets are included in $P_{\mathrm{pre}}$.

For $1\le j\le J_0$, define the ordinary, unweighted reference laws
$\nu_{j,\ell}=\rho_{\theta_{j-1}}P_{u_j}^{\ell}$ for
$0\le\ell\le k_j$. The reverse bound in
\eqref{f:bridge-bounds}, at index $j-1$, and
$L^2(\rho_{\theta_j})$ contraction give
\[
 \left\|\frac{d\nu_{j,\ell}}{d\rho_{\theta_j}}\right\|_{2,\rho_{\theta_j}}^2
 \le1+\chi^2(\rho_{\theta_{j-1}}\Vert\rho_{\theta_j})\le2.
\]
Cauchy--Schwarz and Lemma~\ref{f:moments} therefore imply
\[
 \begin{aligned}
 \mathbb E_{\nu_{j,\ell}}D
 &\le
 \left\|\frac{d\nu_{j,\ell}}{d\rho_{\theta_j}}\right\|_{2,\rho_{\theta_j}}
       \bigl(\mathbb E_{\rho_{\theta_j}}D^2\bigr)^{1/2}\\
 &\le C(1+a\theta_j)\le C'(1+au_j),
 \end{aligned}
\]
using $\theta_j=2u_j/(1+u_j^2)\le2u_j$.
Its expected activation probability is thus
at most $Cv_j$, where
$v_j=(1+Nu_j^2+a u_j)/N$ and $v_j/g_{u_j}=w(u_j)$.
The true unweighted endpoint laws have TV error at most $J_0\,2^{-p}$
by contraction and induction over cooling blocks. Coupling reference
blocks therefore adds at most $\mathcal T\,J_0\,2^{-p}<1$ to the expected total
activation count. Consequently the expectation per path is bounded
by the explicitly computable quantity
\[
 A_{\rm bd}=2^{20}\sum_{j=1}^{J_0} k_jv_j+1
 =\widetilde O(C_{\mathrm{cool}}).
\]
Cap the activations across all $L$ paths at $\lceil128LA_{\rm bd}\rceil$,
returning zero if the cap would be exceeded. Markov's inequality
makes this an additional failure event of probability at most
$1/128$. This argument concerns unweighted path laws solely to bound
work; the AIS second-moment argument above concerns weighted laws.

Every allowed activation will use the sparse Gram implementation below.
Every failed activation costs $O(1)$. Also
$\mathcal T\le N\sum_jk_jv_j=\widetilde O(NC_{\mathrm{cool}})$, because $v_j\ge1/N$.
The single DPP draw per path costs $O(N^3)$. Its sparse Gram
initialization, described in Section~\ref{f:sparse-lift}, costs
$O(K_cn^3)$, where $K_c=M\lceil2s/n\rceil$.

\subsubsection{The initial determinantal sample}
At $u=1$, the collapsed law is a weighted projection determinantal point process (DPP). Set
\[
 \mathsf H_0=\mathcal A\operatorname{diag}(\zeta)\mathcal A^\dagger,\qquad
 \mathsf K=\mathcal A^\dagger \mathsf H_0^{-1}\mathcal A\operatorname{diag}(\zeta).
\]
Its exactly computable normalizer satisfies
$W_1=\det(\mathsf H_0)\prod_js_j$.
The rational matrix $\mathsf K$ is diagonally similar to a Hermitian projection
kernel. If its remaining rank is $k$, choose label $i$ with probability
$\mathsf K_{ii}/k$ and update
$\mathsf K\leftarrow \mathsf K-\mathsf K_{:,i}\mathsf K_{i,:}/\mathsf K_{ii}$. After $r$ choices the selected
set has the desired law. This follows from the determinant Schur
complement formula for conditioning on inclusion; it is the usual
sequential projection-DPP sampler without an eigendecomposition
\cite{LaunayGalerneDesolneux2018}. There is no second-component sample in the collapsed algorithm.
One initial sample costs $O(N^3)$ arithmetic operations.

\subsection{A sparse Gram representation with bounded block size}
\label{f:sparse-lift}

Section~\ref{f:cooling} bounds the number of activated exchanges. Here we bound the cost of each one.

The dense matrix $F$ is convenient for the initial DPP sample. For
the many reinsertion steps, we use a larger matrix with a sparse
structure. Partition the $2s$ ordered events of each macrostep into
$b_c=\lceil2s/n\rceil$ consecutive nonempty chunks, each containing
at most $n$ events. Put
\begin{equation}
 K_c=Mb_c,\qquad d_0=nK_c,\qquad D=d_0+r.
 \label{f:lift-dimension}
\end{equation}
If chunk $a$ contains $k_a$ events, then $1\leq k_a\leq n$.
Let $a(j)$ be the chunk of event $j$. The free propagator before the
first chunk of a macrostep is $T_a=R^2$; before every other chunk it
is $T_a=I_n$. Define the $nK_c\times nK_c$ cyclic matrix
\[
 C_{\mathrm{ch}}=I+\mathcal B_{\mathrm{ch}},\qquad
 (\mathcal B_{\mathrm{ch}})_{a,a+1}=-T_{a+1},\qquad
 (\mathcal B_{\mathrm{ch}})_{K_c,1}=T_1.
\]
For $K_c=1$, the definition means $C_{\mathrm{ch}}=I_n+T_1$.
Let $E_a$ embed $\mathbb C^n$ into physical chunk $a$, and set
\begin{equation}
 \widehat U_{:,j}=E_{a(j)}u_j,\qquad
 \widehat V_{:,j}=E_{a(j)}v_j,\qquad
 (L_{\mathrm{ch}})_{jk}
   =\mathbf1_{\{a(j)=a(k),\ k\leq j\}}v_j^\dagger u_k.
 \label{f:chunk-data}
\end{equation}
The general cyclic identity of Lemma~\ref{d:cyclic-source} gives
\begin{equation}
 F=L_{\mathrm{ch}}-\widehat V^\dagger
                        C_{\mathrm{ch}}^{-1}\widehat U,
 \qquad \det C_{\mathrm{ch}}=\det(I_n+R^{2M})=\det C.
 \label{f:chunk-compression}
\end{equation}
In particular, refining the macrosteps into chunks does not change
$F$. Within a chunk, the inclusive lower-triangular term in
\eqref{f:chunk-data} retains the chronological order of overlapping
channels; no interaction factors are commuted.

Define $d_0$ fixed columns and $2r$ candidate columns in
$\mathbb C^D$ by
\begin{equation}
 \mathcal G=\begin{bmatrix}C_{\mathrm{ch}}\\\widehat V^\dagger\end{bmatrix},
 \qquad a_{c_j}=\begin{bmatrix}0\\e_j\end{bmatrix},\qquad
 a_{b_j}=\begin{bmatrix}\widehat U_{:,j}\\(L_{\mathrm{ch}})_{:,j}\end{bmatrix}.
 \label{f:lift-columns}
\end{equation}
For an $r$-element label set $S$, let $D_S$ contain all the fixed
columns followed by its selected candidate columns in a fixed order.

\begin{lemma}[Lifted determinants and heatbath scores]
\label{f:lift-scores}
For every $r$-element label set $S$,
\[
 |\det D_S|^2=|\det C|^2|\det\mathcal A_{:,S}|^2.
\]
Thus $\mathsf H_S=D_SD_S^\dagger$ is positive definite whenever $S$
is a supported basis. If $x\in S$, solve $\mathsf H_Sz=a_x$.
The unpenalized reinsertion weights are proportional to
\begin{equation}
 w_y=\zeta_y|a_y^\dagger z|^2,
 \qquad y\in\{c_j,b_j:j\in[r]\}.
 \label{f:replacement-scores}
\end{equation}
Every selected $y\neq x$ has score zero and $w_x>0$.
\end{lemma}
\begin{proof}
The Schur complement of $C_{\mathrm{ch}}$ in $D_S$ is the selected
column matrix of $[I_r\mid F]$, by \eqref{f:chunk-compression}.
This proves the determinant formula, including when it vanishes.
If $x$ occupies slot $k$ in $D_S$, then
$a_x^\dagger\mathsf H_S^{-1}=e_k^{\mathsf T}D_S^{-1}$.
Replacing its column by $a_y$ changes the determinant by the factor
$e_k^{\mathsf T}D_S^{-1}a_y$.
The field ratio is $\zeta_y/\zeta_x$; dropping the common denominator
gives \eqref{f:replacement-scores}. The same identity makes all other
selected scores zero and makes the self-reinsertion factor one.
\end{proof}

Group the coordinates of $\mathsf H_S$ by chunk: each block contains
$n$ physical coordinates and the chunk's $k_a$ scalar coordinates.
Its width is at most $2n$. Each fixed column in
\eqref{f:lift-columns} meets only its own chunk and the preceding
chunk. Each candidate column meets only its own chunk, because
$L_{\mathrm{ch}}$ is block diagonal in chunks. Consequently
$\mathsf H_S$ is cyclic block tridiagonal. A basis exchange changes
at most two diagonal chunk blocks; the off-diagonal blocks are fixed.
When $K_c=2$, contributions from both directed edges are included in
the actual Gram blocks.

\subsubsection{A balanced positive-definite factorization}
For $K_c\leq2$ use a direct dense factorization in dimension at most
$4n$. Otherwise take a balanced binary tree of contiguous chunk
intervals. A proper node retains its first and last chunks as boundary,
deduplicating a one-chunk interval, and stores the Schur complement
after eliminating all interior coordinates of that principal interval.
It also stores the elimination multipliers and pivot factorizations.

To merge two adjacent intervals, combine their boundary matrices and
the true edge joining the two intervals. At most four chunks are
present. Eliminate the child boundary chunks that are not endpoints of
the parent; at most two chunks are eliminated. The remaining matrix is
the parent's boundary Schur complement. The edge joining the first and
last chunks of the full cycle is inserted only at the root, before the
root is solved. Every proper interval is a principal submatrix of the
actual positive definite $\mathsf H_S$, and every eliminated pivot is
a positive definite principal-Schur pivot. The procedure therefore
does not need positivity of the full matrix with its cyclic edge
deleted.

A merge costs $O(n^3)$ rational operations. There are $O(K_c)$
nodes, so initialization costs $O(K_cn^3)$ and uses $O(K_cn^2)$
storage. A change to two diagonal blocks affects only their two paths
to the root and costs $O(n^3\log(2K_c))$. All affected blocks are
first replaced by their final values; the affected nodes are then
recomputed from the leaves upward. A positive-score replacement is a
supported final basis, so all these factorizations remain valid. The
singular intermediate Gram matrix obtained by deletion alone is never
factored.

A solve with the stored factors costs $O(K_cn^2)$. In an upward
pass, each node transforms its boundary right-hand sides by its stored
elimination multipliers. After solving the root boundary system with
its cached factors, a downward pass reconstructs the eliminated
variables. Each node acts on vectors of length at most $8n$, using
$O(n^2)$ operations. No dense factorization is repeated during a solve.

\begin{lemma}[Cost of an activation]
\label{f:activation-cost}
After initialization, one activated reinsertion, including its draw and
factorization update, costs
\begin{equation}
 O(K_cn^2+rn+n^3\log(2K_c))
       =\widetilde O(nD+n^3)
 \label{f:sparse-cost}
\end{equation}
rational operations. A clock that does not activate costs $O(1)$.
\end{lemma}
\begin{proof}
Solve $\mathsf H_Sz=a_x$ using the stored factors. Every candidate
column has at most $2n$ entries, so all $2r$ inner products in
\eqref{f:replacement-scores} cost $O(rn)$. An inverse-CDF draw costs
$O(r)$ more. The two changed diagonal blocks are computed in
$O(n^2)$ operations, followed by their tree update. These costs give
\eqref{f:sparse-cost}. Pair occupations and an array of selected labels
allow uniform label choice, activation, and updates of $D(S)$ in
constant work outside activated reinsertions.
\end{proof}

\subsection{Complete algorithm and arithmetic complexity}
\label{f:complexity}

We now combine the annealed importance sampler of Section~\ref{f:cooling} with the sparse updates of Section~\ref{f:sparse-lift} to give Algorithm~\ref{f:algorithm} for estimating $Z_{\mathrm{disc}}$. To bound its total arithmetic work, we first account for matrix setup and path initialization, then combine the bound on the number of activations with the cost of each update.

The matrix setup can use full macrosteps rather than the smaller
chunks. Its $nM$-dimensional cyclic free matrix has $M$ distinct
$n\times n$ inverse blocks up to sign, computed by the finite geometric-series
formula for a cycle with free factor $R^2$. Constructing those blocks
costs $O(Mn^3)$. Each of the $r^2$ entries of $F$ is an endpoint
contraction of such a block, with its triangular correction, and costs
$O(n^2)$. Thus a sufficient setup bound is
\begin{equation}
 T_F=O(Mn^3+r^2n^2).
 \label{f:setup-cost}
\end{equation}
In particular, setup does not invert a dense matrix of dimension
$nK_c$. Compute $|\det C|^2$ and $W_1$ during setup. Each importance
path uses one dense initial DPP sample and initializes the corresponding
sparse Gram tree; these cost $O(r^3+K_cn^3)$ per path.

\begin{algorithm}[htbp]
 \caption{Counting with collapsed importance paths}
 \label{f:algorithm}
 \begin{algorithmic}[1]
 \Require Rational model data, transfer parameters as above, and
          relative accuracy $0<\eta\leq1$.
 \State Remove zero channels.
 \If{$s=0$}
   \State \Return $|\det(I_n+R^{2M})|^2$.
 \EndIf
 \State Normalize endpoints and construct $F$ and $|\det C|^2$.
 \State Construct square weights, $a$, the cooling mesh, mixing counts,
        and $A_{\mathrm{bd}}$.
 \State Compute $W_1$; set $L=\lceil2^{14}\eta^{-2}\rceil$ and the
        fixed-bit draw budget from Appendix~\ref{fa:finite-bits}.
 \For{$\ell=1,\ldots,L$}
   \State Draw $S_0\sim\rho_1$, initialize its Gram tree, and set $Y_\ell=1$.
   \For{$j=0,\ldots,J_0-1$}
     \State $Y_\ell\gets Y_\ell f_j(S_j)$.
     \State Run $k_{j+1}$ clocks at $u_{j+1}$ to obtain $S_{j+1}$,
       using the sparse reinsertion implementation on activation.
     \Statex \hspace{\algorithmicindent}Before each reinsertion, if total activations across all paths would exceed
       $\lceil128LA_{\mathrm{bd}}\rceil$, return zero immediately.
   \EndFor
   \State $Y_\ell\gets Y_\ell\mathbf1_{\{D(S_{J_0})=0\}}$.
 \EndFor
 \State \Return $|\det C|^2W_1L^{-1}\sum_{\ell=1}^L Y_\ell$.
 \end{algorithmic}
\end{algorithm}

\begin{theorem}[Relative counting with explicit arithmetic work]
\label{f:relative-counting}
For the model and transfer parameters above,
Algorithm~\ref{f:algorithm} returns a nonnegative rational estimate
$\widehat Z_{\mathrm{disc}}$ with relative error at most $\eta$ and
probability at least $3/4$. After integer and dyadic preprocessing,
every execution uses
\begin{equation}
 \widetilde O\!\left(\eta^{-2}\left[
 Mn^3+r^2n^2+r^3+K_cn^3+
 (nD+n^3)(Na+a^2)+N(Na+a^2)\right]\right)
 \label{f:structural-work}
\end{equation}
rational arithmetic operations, including matrix setup.
The preprocessing and complete finite-bit implementation have
polynomial bit complexity.
\end{theorem}
\begin{proof}
Lemma~\ref{f:ais-second-moment}, with the bridge bounds and chosen
mixing precision, gives normalized second moment less than four for
an ideal path. Averaging $L$ independent paths gives relative error
$\eta/2$ with statistical failure below $1/64$. Square rounding
contributes its multiplicative $e^{\pm\eta/1000}$ allowance. The
activation cap and fixed-bit coupling contribute failure at most
$1/128$ each. These budgets give the asserted relative accuracy and
success probability.

Each path uses $\widetilde O(Na+a^2)$ activations in expectation,
and the global cap bounds total activations on every execution by
$\widetilde O(\eta^{-2}(Na+a^2))$. There are
$\widetilde O(N(Na+a^2))$ clocks per path. Initialization, matrix
setup, and Lemma~\ref{f:activation-cost} give
\eqref{f:structural-work}. Bridge powers use repeated squaring with
exponents at most $2r$ and $O(\log N)$ rational operations per stage;
this is lower order. Appendix~\ref{fa:finite-bits} proves the finite-bit
claims, including uniform operand bounds along arbitrary supported
trajectories.
\end{proof}

\begin{corollary}[Ground-energy arithmetic bound]
\label{f:ground-complexity}
The ground-energy algorithm may use Algorithm~\ref{f:algorithm} for
its relative counting call. Its counter and transfer construction use
\begin{equation}
 \widetilde O(D^{13/4})
 =\widetilde O\!\left([n(n+s)]^{13/4}
                         \lambda^{13/2}\varepsilon^{-13/2}\right)
 \label{f:final-bound}
\end{equation}
rational operations after polynomial-bit integer and dyadic
preprocessing. The final scalar logarithm has a separate polynomial
bit cost. The output approximates the full-Fock ground energy to
additive error $\varepsilon$ with probability at least $3/4$.
For $s=O(n)$ the arithmetic bound becomes
$\widetilde O(n^{13/2}\lambda^{13/2}\varepsilon^{-13/2})$.
\end{corollary}
\begin{proof}
From \eqref{f:lift-dimension},
$D=nM\lceil2s/n\rceil+2sM=\Theta(M(n+s))$,
$r\leq D$, $N\leq2D$, and $D\geq nM\geq3200n^2$.
Moreover,
\[
 B_{\mathrm{src}}^2\leq(\beta\lambda)^2
       \leq(2/25)nM\leq(2/25)D.
\]
Thus $a=O(D^{3/4})$, $Na+a^2=O(D^{7/4})$, and
$nD+n^3=O(D^{3/2})$. The setup and initialization terms in
\eqref{f:structural-work} are at most $O(D^3)$, and the clock term
is $O(D^{11/4})$. The dominant term at $\eta=1/4$ gives
\[
 \underbrace{\widetilde O(D^{7/4})}_{\text{total activations}}
 \underbrace{\widetilde O(D^{3/2})}_{\text{work per activation}}
 =\widetilde O(D^{13/4}).
\]
Substituting $M=\Theta(n\lambda^2\varepsilon^{-2})$ gives \eqref{f:final-bound}.

Take $\eta=1/4$. If the returned count is nonpositive, return zero
without taking a logarithm. Otherwise approximate its logarithm to
additive accuracy $\beta\varepsilon/16$ and subtract this rational
number divided by $\beta$ from $c$. On the counting success event,
the effective-Hamiltonian and thermal estimates give
\[
 |\widehat E-E_0(H)|\leq
 10\delta\lambda^2+\frac{n\log4}{\beta}
     +\frac{\log(4/3)}{\beta}+\frac{\varepsilon}{16}
 <\varepsilon.
\]
The constant $c$ is restored exactly. No numerical conditioning
assumption or physical spectral gap enters this argument.
\end{proof}

The exponent in \eqref{f:final-bound} is an arithmetic-operation
exponent. The full bit bound additionally includes $P_{\mathrm{pre}}$,
the bit lengths of rational operands, and the scalar logarithm. Its
dependence on $\lambda$ is numerical, so the bound is pseudopolynomial
in that scale. Arbitrarily small positive couplings enter through their
encoding lengths, without a separate reciprocal-coupling parameter.

\section*{AI statement}
The main research ideas and the algorithmic framework were developed by the authors. We used GPT-6-Astra to assist with mathematical derivations and calculations, as well as improving the presentation of the manuscript. All AI-assisted content was carefully checked and substantially revised by the authors. The authors take full responsibility for correctness of the final manuscript.

\section*{Acknowledgments}
We thank Weiming Feng for useful discussions. Ziyi Yang, Yexin Zhang, Yuexin Su, Xinzhao Wang and Tongyang Li are supported by the National Natural Science
Foundation of China under Grant No. 62372006. Xiaoyang Wang is supported by the RIKEN TRIP initiative (RIKEN Quantum) and the UTokyo Quantum Initiative.

\IfFileExists{myhamsplain.bst}
  {\bibliographystyle{myhamsplain}}
  {\bibliographystyle{plain}}
\bibliography{ref}

\newpage

\appendix

\section{Proof details}
\label{app:proof-details}

This appendix gives the two proofs behind the determinantal
representation in Section~\ref{sec:positive}.
Section~\ref{a:transfer-proof} bounds the effective-Hamiltonian error
of the positive transfer, and Section~\ref{a:cyclic-proof} proves the
cyclic source determinant identity.

\subsection{Transfer approximation}
\label{a:transfer-proof}

We prove Proposition~\ref{d:transfer-error}. Set $t=\delta\lambda$,
so $t\leq1/64$. From \eqref{d:strength-bound},
$\delta\norm q/2\leq t/4$, and hence $R\succ0$.
Every interaction factor is invertible by its Neumann series, since
$\gamma_\alpha\norm{B_\alpha}\leq t/2<1$. Thus $X$ is invertible
and $S=XX^\dagger\succ0$.

Let $J=\log G$. The identities of
Lemma~\ref{p:second-quantization-identities} give
\[
 J=d\Gamma(\log R)\otimes I
    +I\otimes d\Gamma(\overline{\log R}).
\]
The logarithmic series and Lemma~\ref{p:spectrum} imply
\begin{equation}
 \begin{split}
 \norm{J+\delta K/2}
 &\leq\frac{\delta^2\norm q_1\norm q}
                  {4(1-\delta\norm q/2)}
 \leq\frac{t^2}{16(1-t/4)}\leq t^2/8,\\
 \norm J&\leq\frac{t/2}{1-t/4}\leq3t/5.
 \end{split}
 \label{a:free-logarithm}
\end{equation}
Indeed $\norm{q^k}_1\leq\norm q^{k-1}\norm q_1$ controls each
term of the matrix logarithm. The exponential remainder
$\norm{e^J-I-J}\leq\frac12e^{\norm J}\norm J^2$ then gives
\begin{equation}
 \norm{G-I+\delta K/2}\leq t^2.
 \label{a:free-linearization}
\end{equation}

For arbitrary matrices $E_0,\ldots,E_s$, expansion in the prescribed
order yields
\[
 \left\|\prod_{\alpha=0}^s(I+E_\alpha)
              -I-\sum_{\alpha=0}^sE_\alpha\right\|
 \leq e^h-1-h,\qquad h=\sum_{\alpha=0}^s\norm{E_\alpha}.
\]
Use $E_0=G-I$ and $E_\alpha=\gamma_\alpha B_\alpha$.
Equations~\eqref{d:strength-bound} and
\eqref{a:free-linearization} give
$h\leq t+t^2\leq3t/2$, whence $e^h-1-h\leq2t^2$.
Putting
\[
 D=-K/2+\tfrac12\sum_\alpha g_\alpha B_\alpha,
 \qquad \norm D\leq\lambda/2,
\]
we obtain $X=I+\delta D+E$ with $\norm E\leq3t^2$.
Since $D+D^\dagger=-H_0$,
\begin{equation}
 \norm{S-I+\delta H_0}
 \leq6t^2+(t/2+3t^2)^2\leq7t^2,
 \qquad \norm{S-I}\leq t+7t^2.
 \label{a:positive-linearization}
\end{equation}
The last bound is at most $71/4096<1$, so the logarithmic series gives
\[
 \norm{\log S-(S-I)}
 \leq\frac{\norm{S-I}^2}{2(1-\norm{S-I})}\leq t^2.
\]
Combining this with \eqref{a:positive-linearization} proves
$\norm{-\delta^{-1}\log S-H_0}\leq8\delta\lambda^2$, and in
particular the asserted bound with constant $10$. Finally,
$S^M=e^{-\beta H_{\mathrm{eff}}}$ follows directly from spectral
calculus. No step of this proof reorders interaction factors or
assumes they commute.

\subsection{Cyclic determinant identity}
\label{a:cyclic-proof}

We prove Lemma~\ref{d:cyclic-source}. In this subsection, a chunk is
any consecutive group of events contained in one macrostep. For
chunk $a$, let
\[
 P_a(z)=\prod_{j:\,a(j)=a}(I_n+z_ju_jv_j^\dagger),\qquad
 Z=\diag(z_1,\ldots,z_r).
\]
The trace--determinant identity in
\eqref{d:source-product-determinant} reads
$f(z)=\det(I_n+T_1P_1\cdots T_kP_k)$.
The cyclic block elimination used below also holds for $k=1$, in
which case it is the elementary identity $I_n+T_1P_1$.

Consider the bordered block matrix
\begin{equation}
 \mathcal M(z)=
 \begin{pmatrix}
  C_k&\widehat U Z\\
  \widehat V^\dagger&I_r+LZ
 \end{pmatrix}.
 \label{a:bordered-source}
\end{equation}
First eliminate its upper-left block. This gives
\begin{equation}
 \det\mathcal M(z)
   =\det C_k\det\!\left(I_r+
           (L-\widehat V^\dagger C_k^{-1}\widehat U)Z\right).
 \label{a:upper-elimination}
\end{equation}
Invertibility of $C_k$ will follow from the free case below.

For the second elimination, initially take values of $z$ for which
all the factors $I_n+z_ju_jv_j^\dagger$ are invertible. The resulting
identities are rational identities and hence extend polynomially to
all $z$. Within a chunk, write $U_a,V_a$ for its endpoint matrices,
$L_a$ for its block of $L$, and $Z_a$ for its variables. Since $L_a$
is lower triangular with diagonal $v_j^\dagger u_j$,
\begin{equation}
 \det(I+L_aZ_a)=\prod_{j:\,a(j)=a}
                 (1+z_jv_j^\dagger u_j)=\det P_a.
 \label{a:local-source-determinant}
\end{equation}
Sequential rank-one inversion gives
\begin{equation}
 I_n-U_aZ_a(I+L_aZ_a)^{-1}V_a^\dagger=P_a^{-1}.
 \label{a:local-source-inverse}
\end{equation}
For clarity, this identity can also be obtained by forward substitution.
If $(I+L_aZ_a)y=V_a^\dagger x$, then its $j$th equation implies
\[
 y_j=\frac{v_j^\dagger(x-\sum_{\ell<j}z_\ell u_\ell y_\ell)}
                  {1+z_jv_j^\dagger u_j}.
\]
Starting with $x$, subtracting $z_ju_jy_j$ at step $j$ therefore
applies $(I_n+z_ju_jv_j^\dagger)^{-1}$ to the current vector.
After all events the result is
$P_a^{-1}x$, proving \eqref{a:local-source-inverse} with the
specified order.

Eliminating the lower-right block in \eqref{a:bordered-source},
using \eqref{a:local-source-determinant} and
\eqref{a:local-source-inverse}, now gives
\[
 \begin{split}
 \det\mathcal M(z)
 &=\left(\prod_a\det P_a\right)
       \det\!\left(\diag(P_1^{-1},\ldots,P_k^{-1})
                        +\mathcal B_k\right)\\
 &=\det\!\left(I_{nk}+\mathcal B_k
                       \diag(P_1,\ldots,P_k)\right)\\
 &=\det(I_n+T_1P_1\cdots T_kP_k)=f(z).
 \end{split}
\]
The last equality follows by eliminating the first $k-1$ block rows
of the cyclic matrix: the remaining Schur complement is
$I_n+T_1P_1\cdots T_kP_k$. This also checks the signs in
\eqref{d:cyclic-matrix}. At $z=0$ the same calculation gives
$\det C_k=\det(I_n+T_1\cdots T_k)=\det(I_n+R^{2M})>0$,
which justifies the upper-left elimination.
Equation~\eqref{a:upper-elimination} proves the determinant formula
for $F$ and its principal-minor coefficients.

To verify that the whole matrix $F$, rather than just its principal
minors, is unchanged by chunking, let $m_a$ be the macrostep containing
chunk $a$. All free propagators commute, since each is $I_n$ or $R^2$.
Direct substitution in the cyclic recurrence gives
\[
 E_a^\dagger C_k^{-1}E_b=(I_n+Q_R)^{-1}
 \begin{cases}
 R^{2(m_b-m_a)},&a\leq b,\\
 -R^{2(M+m_b-m_a)},&a>b.
 \end{cases}
\]
Contract with $v_j^\dagger,u_\ell$ and subtract this block from
$L_{j\ell}$. For different chunks this gives the two chronological
branches of \eqref{d:chronological-kernel}. Within the same chunk,
it gives $v_j^\dagger P_Ru_\ell$ for $\ell\leq j$ and
$v_j^\dagger(P_R-I_n)u_\ell$ for $j<\ell$.
Thus every partition gives exactly
\eqref{d:chronological-kernel}, proving partition independence.
For $k=1$ the formula is $C_1^{-1}=(I_n+T_1)^{-1}$;
for $k=2$ the two directed cyclic blocks are distinct and both
retain their prescribed signs.

\section{Algorithm details}
\label{app:algorithm-details}


In this section, we provide the lemmas and implementation details used to analyze the ground-energy algorithms of Sections~\ref{sec:counting} and~\ref{sec:fast}. Appendix~\ref{a:logarithm-section} bounds the bit cost of computing the logarithm that both algorithms use to convert a partition function estimate into a ground-energy estimate. For the faster algorithm of Section~\ref{sec:fast}, Appendix~\ref{fa:wick} proves the ordered insertion formula used in Lemma~\ref{f:source-identity} to express the deformed minor sum as a positive trace, including when orbital channels overlap. Appendix~\ref{fa:finite-bits} establishes the polynomial bit complexity of the faster counter by analyzing weight rounding, finite-bit random draws, and the lengths of rational operands.

\subsection{The scalar logarithm in the bit model}
\label{a:logarithm-section}

\begin{lemma}[Rational scalar logarithm]
 \label{a:scalar-logarithm}
 Given a positive rational $z$ of binary encoding length $b_z$ and
 rational $\tau>0$ of encoding length $b_\tau$, a deterministic algorithm returns a rational
 $\ell$ with $|\ell-\log z|\leq\tau$ in bit time polynomial in
 $b_z,b_\tau$ and $\log(2+\tau^{-1})$.
\end{lemma}
\begin{proof}
 Exact integer comparisons give $z=2^ky$ with $1\leq y<2$ and
 $|k|=O(b_z)$. For $1\leq y\leq2$, put $x=(y-1)/(y+1)$, so
 $0\leq x\leq1/3$. The identity
 \[
  \log y=2\sum_{j=0}^L\frac{x^{2j+1}}{2j+1}+E_L,
  \qquad
  0\leq E_L\leq
       \frac{2x^{2L+3}}{(2L+3)(1-x^2)}
 \]
 follows by integrating the geometric series for $(1-x^2)^{-1}$.
 The same formula at $x=1/3$ computes $\log2$.
 Choose $L=O(\log(2+(|k|+1)/\tau))$ so that each error is at most
 $\tau/(2(|k|+1))$, and add the approximations to
 $k\log2+\log y$. All partial sums are rational and have encoding
 length polynomial in $b_z,L$. Exact rational arithmetic therefore
 has the stated bit bound.
\end{proof}

This final scalar cost is separate from the arithmetic operation
bound for the specialized counter. The complete bit running time
also includes the input encoding length and the bit lengths of all
rational operands.

\subsection{A finite CAR proof for the pair source}
\label{fa:wick}

We prove the ordered insertion formula used in
Lemma~\ref{f:source-identity}, including nonorthogonal endpoints at
the same macro time. All calculations take place in the finite
$2^n$-dimensional one-component Fock space. For a column $u$ and a row
$v$, define
\[
 \mathsf C(u)=\sum_i u_i a_i^\dagger,\qquad
 \mathsf A(v)=\sum_i v_i a_i.
\]
The row convention in $\mathsf A(v)$ has no extra conjugation;
in particular $a(v)=\mathsf A(v^\dagger)$ when $v$ is a column.
Put
\[
 Q_R=R^{2M},\qquad P_R=Q_R(I_n+Q_R)^{-1},\qquad
 \omega(Y)=\frac{\Tr(\Gamma(Q_R)Y)}{\det(I_n+Q_R)}.
\]
Since $R\succ0$, all inverses here exist. The identities for second
quantization give
\[
 \Gamma(Q_R)\mathsf C(u)\Gamma(Q_R)^{-1}=\mathsf C(Q_Ru),\qquad
 \Gamma(Q_R)\mathsf A(v)\Gamma(Q_R)^{-1}=\mathsf A(vQ_R^{-1}).
\]

Let $W=B_2\cdots B_{2k}$ be a word of linear creation and
annihilation operators. Anticommuting a creation operator through
the word and then using trace cyclicity gives
\begin{equation}
 \begin{split}
 &\omega(\mathsf C(u)W)+\omega(\mathsf C(Q_R^{-1}u)W)\\
 &\quad=\sum_{j=2}^{2k}(-1)^j
       \{\mathsf C(u),B_j\}
       \omega(B_2\cdots\widehat B_j\cdots B_{2k}).
 \end{split}
 \label{fa:creation-recursion}
\end{equation}
The anticommutators are scalar. Replacing $u$ on the right by
$P_Ru$ makes the sum of the two creation operators on the left
equal $\mathsf C(u)$, because $(I_n+Q_R^{-1})P_R=I_n$.
An annihilation-first recursion is obtained in the same way by
replacing its row $v$ by $v(I_n+Q_R)^{-1}$. For a two-letter word
the contractions are therefore
\begin{equation}
 \omega(\mathsf C(u)\mathsf A(v))=vP_Ru,\qquad
 \omega(\mathsf A(v)\mathsf C(u))=v(I_n-P_R)u.
 \label{fa:two-point}
\end{equation}
Repeated use of these recursions expresses an even word as the signed
sum of its pair contractions. Only creation--annihilation pairs have
nonzero contraction. Expanding along the first creation or annihilation
operator is the Laplace expansion of the associated chronological
determinant. This proves the determinant formula by induction on the
number of pairs. It is also the finite-dimensional special case of
the chronological Wick rule in \cite[Lemma~3.1]{BruPedra2017}.
Words with unequal creation and annihilation counts have zero trace:
conjugation by $e^{it\mathsf N}$, where
$\mathsf N=\sum_i a_i^\dagger a_i$, leaves $\Gamma(Q_R)$ invariant
and multiplies such a word by a nonconstant phase.

For an event $j$ in macrostep $m(j)\in\{0,\ldots,M-1\}$, move
all free factors $\Gamma(R^2)$ to the left of the ordered word. The
event's remaining endpoints are
\[
 \mathsf C(R^{-t_j}u_j),\qquad
 \mathsf A(v_j^\dagger R^{t_j}),\qquad t_j=2(M-1-m(j)).
\]
If both are present at the same event, the creation endpoint precedes
the annihilation endpoint. Let $I$ be the selected creation indices
and $J$ the selected annihilation indices. The chronological determinant
has row index $j\in J$, column index $k\in I$, and entry
\begin{equation}
 v_j^\dagger R^{t_j}
          (P_R-\mathbf1_{\{j<k\}}I_n)R^{-t_k}u_k.
 \label{fa:chronological-contraction}
\end{equation}
When the annihilation endpoint precedes the creation endpoint, the
two-point contraction $I_n-P_R$ acquires the sign from the determinant
ordering, giving $P_R-I_n$. Otherwise the contraction is $P_R$.
Thus \eqref{fa:chronological-contraction} is exactly the matrix entry
$F_{jk}$ in \eqref{d:chronological-kernel}. In particular, the term
$-v_j^\dagger u_k$ cannot be discarded for distinct nonorthogonal
events at the same macro time.

We have proved that the one-component insertion trace is
\[
 \epsilon_{I,J}\det C\det F[J,I]\quad\text{if }|I|=|J|,
\]
and zero otherwise. Its sign depends only on the ordering of endpoints.
In the auxiliary tensor source, both components choose exactly the same
endpoints. The second trace is the complex conjugate of the first,
so their product is $|\det C|^2|\det F[J,I]|^2$.
At an adjoint event, the two endpoints are swapped and the event
order is reversed, as in the adjoint product. Multiplying the scalar
endpoint weights and summing gives \eqref{f:source-trace}.

\subsection{Finite-bit implementation of the faster counter}
\label{fa:finite-bits}

\subsubsection{Square weights and deterministic control parameters}
For a positive rational $\gamma_\alpha=P_\alpha/Q_\alpha$, with
positive integers $P_\alpha,Q_\alpha$, put
\[
 k_\alpha=\ell_2(Q_\alpha)+\ell_2(10^6N/\eta),\qquad
 s_\alpha=2^{-k_\alpha}
       \left\lfloor\sqrt{\left\lfloor
                      2^{2k_\alpha}\gamma_\alpha
                       \right\rfloor}\right\rfloor.
\]
Integer division and integer square root compute this rational number.
The identity $\lfloor\sqrt{\lfloor x\rfloor}\rfloor=
\lfloor\sqrt x\rfloor$ shows that it is a downward approximation
to $\sqrt{\gamma_\alpha}$ with error less than $2^{-k_\alpha}$.
Since $\gamma_\alpha\leq1/4$ and $P_\alpha\geq1$, one has
$\sqrt{\gamma_\alpha}\geq1/Q_\alpha$. This proves
\eqref{f:square-accuracy}, with room in its constant, and also shows
that $s_\alpha>0$.

The integer $k_\alpha$ has size
$O(\log Q_\alpha+\log N+\log\eta^{-1})$. The shifted integer and
its square root consequently have polynomial encoding length.
Long division and integer square root by binary search have polynomial
bit cost. The endpoint normalization exponents likewise follow from
exact comparisons of rational squared norms with powers of four.
These operations use binary encodings; they are included in
$P_{\mathrm{pre}}$, rather than charged as unit-cost rational
arithmetic.

The same convention covers $a$, $M$, the dyadic cooling steps, and
the mixing counts, path count, activation cap, and fair-bit precision.
The cooling mesh has polynomially many stages. Each endpoint $u_j$
is a product of at most $J_0$ rational factors with polynomial encoding
length, so its exact encoding length is polynomial as well. Exact
integer comparisons determine every ceiling or dyadic exponent.
Thus all deterministic control data have polynomial bit complexity
in $n,s,M,\eta^{-1}$ and the original input encoding length. Small
positive strengths contribute their encoding lengths, with no
additional reciprocal-strength parameter.

\subsubsection{Fixed-bit random draws and preservation of support}
The ideal analysis uses exact categorical and Bernoulli probabilities.
Implement each draw with a fixed budget
\begin{equation}
 b_{\mathrm{draw}}=\ell_2\!\left(
       2^{20}LN(\mathcal T+N)/\xi\right),\qquad \xi=1/128.
 \label{fa:draw-precision}
\end{equation}
Take a uniform $b_{\mathrm{draw}}$-bit grid point in $[0,1)$ and
choose the first cumulative probability strictly greater than that
point. All cumulative probabilities are computed as exact rationals.
Zero-length intervals are skipped. In particular, a zero replacement
score is never selected, so every maintained basis is supported.

A categorical draw with at most $N$ alternatives differs from its ideal
continuous-uniform inverse-CDF draw only in grid cells meeting one of
its cumulative-probability boundaries. Its coupling error is at most
$N2^{-b_{\mathrm{draw}}}$. There are at most
$L(3\mathcal T+N)$ draws, up to an absolute constant: the initial
DPP choices, one label and activation draw per clock, and one
reinsertion draw per activation. Conditioning successively on their
common past and using a union bound, \eqref{fa:draw-precision}
couples the complete implemented run to the ideal capped run with
failure at most $1/128$.
On coupling agreement the activation cap agrees as well. The success
analysis therefore uses the ideal path variance, the cap failure, and
this trajectory coupling. No separate spectral assertion about the
implemented finite-grid transition rows is needed.

\subsubsection{Uniform operand lengths and complete bit complexity}
Every fixed or candidate column in \eqref{f:lift-columns} is Gaussian
rational with polynomial encoding length. Each current Gram matrix
$\mathsf H_S$ is a sum of $D$ outer products of such fixed columns.
Its entries thus have a polynomial encoding bound uniform over all
bases $S$. Clear their denominators and apply Hadamard's inequality:
all minors have polynomial bit length. On a supported basis, entries
of its inverse and solutions are ratios of these minors. Each stored
Schur complement and elimination multiplier is likewise a ratio of
minors of a principal submatrix, or a product of a bounded number of
such ratios. When fractions are reduced, their lengths remain
polynomial independently of the number of preceding exchanges.

The same argument applies to the exact initial DPP kernel and its
conditional updates. Entries of its matrix are input-bounded rational
data, and conditional kernels are Schur complements. All positive
sampling probabilities and cumulative sums therefore have polynomial
encoding length, including arbitrarily small nonzero probabilities.
The numerical conditioning of a matrix does not enter this exact
rational implementation.

Each bridge is a rational power with exponent at most $2r$, computed
by repeated squaring. A path weight is a product of $J_0$ such terms;
its encoding length is polynomial in the mesh and input sizes. Taking
the mean over $L$ rational path weights also has polynomial encoding
length. Abort outputs have fixed finite encoding. Hence the runtime
bound holds on every execution, including all failure events.

Let $A_{\mathrm{rat}}$ denote \eqref{f:structural-work}, and let
$\mathcal B$ bound the bit lengths of all integer operands in the
rational phase. Ordinary exact arithmetic and fraction reduction give
the conservative counter bit bound
\begin{equation}
 P_{\mathrm{pre}}+\widetilde O(\mathcal B^2A_{\mathrm{rat}}).
 \label{fa:bit-bound}
\end{equation}
The rational matrix setup is included in $A_{\mathrm{rat}}$ and is
distinct from the binary preprocessing cost $P_{\mathrm{pre}}$.
The final scalar logarithm is computed separately using
Lemma~\ref{a:scalar-logarithm}. Substitution of the ground-energy
parameters makes the complete bit cost polynomial in
$n,s,\lambda,\varepsilon^{-1}$ and input encoding length.
The arithmetic exponent $13/2$ when $s=O(n)$ is not an encoding-length
independent bit exponent.

\section{Examples separating the two Hamiltonian classes}
\label{sec:comparison-examples}

We compare Definition~\ref{p:model-definition} with the attractive
Kramers class of Rayudu, Takahashi, Parekh, and Zhao
\cite[Definition~1]{RayuduEtAl2026}. The latter consists of
\begin{equation}
 H_{\mathrm K}=d\Gamma(h)-\sum_{\alpha}U_\alpha
                 n(v_\alpha)n(\Theta v_\alpha),\qquad
 h=h^\dagger,\quad \Theta h\Theta^{-1}=h,\quad
 \Theta^2=-I,\quad U_\alpha\geq0,
 \label{intro:kramers-form}
\end{equation}
where $\Theta$ is antiunitary and the $v_\alpha$ are unit vectors.
Here $n(v)=c^\dagger(v)c(v)$, with $c$ denoting annihilation operators
on the full one-particle space. Kramers partners are orthogonal:
$\langle v,\Theta v\rangle=0$. Both examples below use four physical
modes. The separations persist under linear canonical changes of
these same modes, including particle--hole transformations, and
under scalar shifts. They make no claim about reductions using
additional modes, projections, or general computational gadgets.

\begin{proposition}[A conjugate-pair interaction outside the Kramers density class]
 \label{intro:pair-separation}
 Let $\Delta_i=c_{i\downarrow}c_{i\uparrow}$, for $i=1,2$, and $J>0$.
 The pair-hopping Hamiltonian
 \begin{equation}
 H_{\mathrm{pair}}=-J(\Delta_1^\dagger\Delta_2+
                            \Delta_2^\dagger\Delta_1)
 \label{intro:pair-hopping}
 \end{equation}
 has the form of Definition~\ref{p:model-definition}, but no linear
 canonical change of its four modes transforms it, even up to a
 scalar, into \eqref{intro:kramers-form}.
\end{proposition}
\begin{proof}
 In our model take $q=0$, $c=0$, $s=1$, $u_1=e_1$, $v_1=e_2$, and
 $g_1=2J$. The parity convention \eqref{p:parity-convention} gives
 $A_1\otimes\overline{A_1}=\Delta_1^\dagger\Delta_2$.
 The operator $H_{\mathrm{pair}}$ has simple eigenvalues $-J,+J$
 on the span of the two onsite-pair states and vanishes on the
 remaining fourteen states. In particular, it is zero on odd parity.

 We prove the stronger obstruction to any representation
 $K=cI+d\Gamma(h)-V$, where
 $V=\sum_\alpha t_\alpha n(v_\alpha)n(w_\alpha)$,
 $t_\alpha\geq0$, and each pair $v_\alpha,w_\alpha$ is orthonormal.
 Different pairs may overlap. Put $s_V=\sum_\alpha t_\alpha$.
 The traces of one pair-density projection on the sectors of particle
 numbers $0,1,2,3,4$ are respectively $0,0,1,2,1$.

 Suppose first that $K$ is unitarily equivalent to $H_{\mathrm{pair}}$
 by a parity-preserving unitary. It is zero on odd parity, so its
 one-particle restriction gives $h=-cI_4$. The three-particle trace
 then gives $s_V=-4c$. Both the vacuum and fully occupied state have
 energy $c$, since $c+\Tr h-s_V=c$. If $c\ne0$, this contradicts
 simplicity of every nonzero eigenvalue of $H_{\mathrm{pair}}$.
 If $c=0$, then $s_V=0$, whence $V=0$ and $h=0$, also a contradiction.

 If the unitary reverses parity, $K$ is zero on even parity. The
 vacuum gives $c=0$; the fully occupied state and the two-particle
 trace give $\Tr h-s_V=0$ and $3\Tr h-s_V=0$. Hence $s_V=0$ and
 $V=0$. Vanishing of $d\Gamma(h)$ on the two-particle sector says
 that every pair sum of the four eigenvalues of $h$ is zero, forcing
 $h=0$. This again contradicts $H_{\mathrm{pair}}\ne0$.

 A linear canonical change sends Majorana operators to a real
 orthogonal linear combination. Their product, proportional to
 parity, changes by the determinant of that orthogonal matrix.
 Its implementing unitary therefore preserves or reverses parity.
 Both cases have been excluded, and Kramers density attractions
 are among the orthogonal-pair interactions just considered.
\end{proof}

The reverse separation needs more than an off-diagonal component block
in a hopping matrix: a Kramers-symmetric Hermitian matrix alone can
always be diagonalized into degenerate pairs. The obstruction below
concerns the quadratic conserved quantities of the full interacting
Hamiltonian and therefore also excludes a hidden change of component basis.

\begin{proposition}[A Kramers density Hamiltonian without a conjugate component decomposition]
 \label{intro:kramers-separation}
 On four modes let $\mathsf N=\sum_{j=1}^4c_j^\dagger c_j$,
 $\Theta v=J_4\bar v$, and
 \[
 J_4=\begin{pmatrix}0&-I_2\\I_2&0\end{pmatrix},\qquad
 (v_1,\ldots,v_6)=
 (e_1,e_2,e_1+e_2,e_1+ie_2,e_1+e_4,e_1+ie_4),
 \]
 with $(g_1,\ldots,g_6)=(1,2,3,5,7,11)$. Then
 \begin{equation}
 H_{\mathrm{irr}}=64\mathsf N-
             \sum_{\alpha=1}^6g_\alpha
                    n(v_\alpha)n(\Theta v_\alpha)
 \label{intro:irreducible-kramers}
 \end{equation}
 belongs to the attractive Kramers class after normalizing its
 endpoint vectors. It admits no representation of
 Definition~\ref{p:model-definition}, even after a linear canonical
 change of these same four modes and a scalar shift.
\end{proposition}
\begin{proof}
 The matrix $64I_4$ commutes with $\Theta$ and $\Theta^2=-I_4$.
 Normalizing $v_\alpha$ multiplies its coupling by
 $\norm{v_\alpha}_2^4$, giving \eqref{intro:kramers-form}.
 All coefficients of the Hamiltonian itself are Gaussian rational.

 On the two-particle sector $\F_4^{(2)}$, use the occupation states
 $\ket{\{i,j\}}=c_i^\dagger c_j^\dagger\ket{\varnothing}$ for
 $i<j$. Define the pair states
 \[
 \ket{\psi_\alpha}
   =c^\dagger(v_\alpha)c^\dagger(\Theta v_\alpha)\ket{\varnothing}
   =\sum_{1\leq i<j\leq4}
     \bigl[(v_\alpha)_i(\Theta v_\alpha)_j
           -(v_\alpha)_j(\Theta v_\alpha)_i\bigr]\ket{\{i,j\}}.
 \]
 The interaction in $H_{\mathrm{irr}}$ restricts to $-W$, where
 $W=\sum_{\alpha=1}^6g_\alpha\ket{\psi_\alpha}\bra{\psi_\alpha}$.
 In the ordered occupation basis
 \[
 (\ket{\{1,2\}},\ket{\{1,3\}},\ket{\{1,4\}},
  \ket{\{2,3\}},\ket{\{2,4\}},\ket{\{3,4\}}),
 \]
 it is the matrix
 \[
 W=\begin{pmatrix}
 18&-7+11i&0&0&-7+11i&-4\\
 -7-11i&27&3+5i&3-5i&26&-7+11i\\
 0&3-5i&8&-2&3-5i&0\\
 0&3+5i&-2&8&3+5i&0\\
 -7-11i&26&3+5i&3-5i&28&-7+11i\\
 -4&-7-11i&0&0&-7-11i&18
 \end{pmatrix}.
 \]
 Since $W\succeq0$ and $\Tr W=107$, the restriction
 $\left.H_{\mathrm{irr}}\right|_{\F_4^{(2)}}=128I_6-W$
 is positive definite.

 We show that every Hermitian quadratic operator commuting with
 $H_{\mathrm{irr}}$ belongs to
 $\operatorname{span}_{\mathbb R}\{I,\mathsf N\}$.
 Write such an operator as $Q=cI+d\Gamma(X)+P^\dagger+P$,
 where $X=X^\dagger$ and $P^\dagger$ is a pair-creation operator.
 Applying $[H_{\mathrm{irr}},Q]=0$ to the vacuum gives
 $(128I_6-W)P^\dagger\ket{\varnothing}=0$.
 Thus $P^\dagger=0$ and $P=0$.
 Let $X^{[2]}$ be the matrix of $d\Gamma(X)$ on $\F_4^{(2)}$ in
 the occupation basis above. For $i<j$ and $k<\ell$, its entries are
 \[
 (X^{[2]})_{\{i,j\},\{k,\ell\}}
   =X_{ik}\delta_{j\ell}-X_{i\ell}\delta_{jk}
       -X_{jk}\delta_{i\ell}+X_{j\ell}\delta_{ik}.
 \]
 The remaining two-particle commutation relation is therefore
 \begin{equation}
 [X^{[2]},W]=0.
 \label{intro:commutant-equation}
 \end{equation}

 Here the only solutions of \eqref{intro:commutant-equation}, even
 over $\mathbb C$, are scalar matrices. To give an exact reproducible
 certificate, form the $36\times16$ matrix $C$ whose $(i,j)$ column
 is the row-major vectorization of $[(E_{ij})^{[2]},W]$, with the
 sixteen pairs $(i,j)$ also in row-major order. The determinant of
 its first fifteen rows and first fifteen columns is
 \[
 -731185834229760-7626940877537280i\ne0.
 \]
 Hence $\operatorname{rank}C\geq15$. The identity matrix belongs to
 its kernel, so $\operatorname{rank}C=15$ and
 $\ker C=\mathbb C I_4$. This proves the assertion about quadratic
 conserved quantities.

 Every Hamiltonian of Definition~\ref{p:model-definition} commutes
 with the two component-number operators $\mathsf N_\uparrow$ and
 $\mathsf N_\downarrow$. These are linearly independent modulo
 constants. Conjugation by a linear canonical transformation preserves
 quadratic operators and their independence modulo constants.
 Their pullbacks would therefore give two independent commuting
 quadratic operators modulo constants for $H_{\mathrm{irr}}$.
 The preceding calculation leaves only the one-dimensional span of
 $\mathsf N$ modulo constants, a contradiction. Scalar shifts do
 not change any commutator.
\end{proof}

\section{A common extension by Kramers pair transfers}
\label{sec:common-extension}

This appendix combines the quaternionic minor polynomial of
Rayudu, Takahashi, Parekh, and Zhao~\cite[Lemma~8]{RayuduEtAl2026}
with the full Chen--Liu counting theorem. It gives a polynomial-time
algorithm for a common extension of their attractive Kramers density
class and our conjugate-pair orbital interaction class. The extension is a
consequence of these counting tools and the determinant reduction below;
we do not claim a new quaternionic log-concavity theorem or the
specialized arithmetic exponent of Section~\ref{sec:fast}.

\subsection{The class and its computational guarantee}

For an integer $N\geq1$, work on the full Fock space of $2N$ modes, in a supplied
Kramers-adapted one-particle basis. In this basis fix
\[
 \Theta z=J_N\overline z,\qquad
 J_N=\begin{pmatrix}0&-I_N\\I_N&0\end{pmatrix},
 \qquad \Theta^2=-I_{2N}.
\]
For $v\in\mathbb C^{2N}$ define
\[
 \Delta(v)=a(\Theta v)a(v),\qquad
 \Delta^\dagger(v)=a^\dagger(v)a^\dagger(\Theta v).
\]
The orthogonality of Kramers partners gives
$\norm{\Delta(v)}=\norm v_2^2$.
Consider the Hamiltonians
\begin{equation}
 \begin{split}
 H={}&cI+d\Gamma(h)
 -\frac12\sum_{\alpha=1}^s g_\alpha
 \bigl[\Delta^\dagger(u_\alpha)\Delta(v_\alpha)
       +\Delta^\dagger(v_\alpha)\Delta(u_\alpha)\bigr],\\
 &h=h^\dagger,\qquad hJ_N=J_N\overline h,\qquad g_\alpha\geq0.
 \end{split}
 \label{g:model}
\end{equation}
The supplied vectors may overlap, have arbitrary norms, and have
unequal creation and annihilation endpoints. All terms preserve total
particle number; separate component particle numbers are not required.

For the bit-model statement, $c,g_\alpha$ and the real and imaginary
parts of $h,u_\alpha,v_\alpha$ are rational in the supplied basis.
The structural decomposition in \eqref{g:model} is part of the input.
We do not address finding a hidden Kramers basis or recognizing an
arbitrary quartic Hamiltonian as a member of this class. Put
\begin{equation}
 L=1+\sum_{i,j}(|\operatorname{Re}h_{ij}|+
                       |\operatorname{Im}h_{ij}|)
       +\sum_\alpha g_\alpha\norm{u_\alpha}_2^2\norm{v_\alpha}_2^2.
 \label{g:strength}
\end{equation}
Then $\norm{H-cI}\leq L-1$. Let $b$ denote the full input encoding
length, including rational $0<\varepsilon\leq1$.

\begin{theorem}[Ground energy for Kramers pair-transfer Hamiltonians]
 \label{g:main}
 For the supplied rational data in \eqref{g:model}, a classical
 randomized algorithm returns a rational $\widehat E$ satisfying
 \[
 \Pr\bigl[|\widehat E-E_0(H)|\leq\varepsilon\bigr]\geq3/4
 \]
 in bit time $\operatorname{poly}(N,s,L,\varepsilon^{-1},b)$ on every
 execution. The minimum is over the full Fock space and requires no
 physical spectral gap. Dependence on $L$ is numerical; the scalar
 shift $c$ is added exactly. No $13/2$ arithmetic exponent is asserted.
\end{theorem}

At the level of Hamiltonian forms, \eqref{g:model} contains both
our model and the model in~\cite{RayuduEtAl2026}. If $u_\alpha=v_\alpha$,
orthogonality gives
$\Delta^\dagger(v_\alpha)\Delta(v_\alpha)
  =n(v_\alpha)n(\Theta v_\alpha)$, recovering the attractive Kramers
density interactions. To recover our main model, set
$h=q\oplus\overline q$ and embed every endpoint into the first component,
$u\mapsto(u,0)$ and $v\mapsto(v,0)$. The parity convention
\eqref{p:parity-convention} then gives
\[
 \Delta^\dagger((u,0))\Delta((v,0))
    =a^\dagger(u)a(v)\otimes
                 \overline{a^\dagger(u)a(v)}.
\]
The computational inclusion applies to representatives whose data
have the stated rational encoding in the supplied adapted basis.

\subsection{A positive transfer and its pair determinant expansion}

Write $K=d\Gamma(h)$ and
$B_\alpha=\Delta^\dagger(u_\alpha)\Delta(v_\alpha)$. For rational
$\delta>0$ with $t:=\delta L\leq1/64$, set
\begin{equation}
 R=I_{2N}-\delta h/2,\qquad G=\Gamma(R),\qquad
 X=G\prod_{\alpha=1}^s(I+\delta g_\alpha B_\alpha/2),
 \qquad S=XX^\dagger.
 \label{g:transfer}
\end{equation}
Here $R\succ0$ commutes with $\Theta$. The interaction factors are
invertible by their Neumann series, so $S\succ0$. The error argument
of Section~\ref{a:transfer-proof} applies with the following free
logarithm estimates:
\[
 \norm{\log G+\delta K/2}
 \leq\frac{\delta^2\norm h_1\norm h}
                 {8(1-\delta\norm h/2)}
 \leq\frac{t^2}{8(1-t/2)},\qquad
 \norm{\log G}\leq\frac{t}{2(1-t/2)}.
\]
Consequently $G=I-\delta K/2+E_G$ with $\norm{E_G}\leq t^2$.
The ordered-product expansion gives
$X=I+\delta D+E$, with
$D=-K/2+\frac12\sum_\alpha g_\alpha B_\alpha$,
$\norm D\leq L/2$, and $\norm E\leq3t^2$.
Thus $\norm{S-I+\delta(H-cI)}\leq7t^2$ and
$\norm{S-I}\leq t+7t^2$. The logarithmic remainder is at most $t^2$,
proving
\begin{equation}
 \norm{-\delta^{-1}\log S-(H-cI)}\leq10\delta L^2.
 \label{g:transfer-error}
\end{equation}
No interaction factors have been commuted in this estimate.

Discard zero channels. For $s\geq1$, expand $\Tr S^M$ into
$r=2sM\geq2$ ordered events. At an adjoint event interchange its
endpoints. Write $B_j=\Delta^\dagger(u_j)\Delta(v_j)$ and
$\gamma_j=\delta g_\alpha/2$ for an event of channel $\alpha$.
Trace cyclicity combines free factors to $\Gamma(R^2)$ between
macrosteps. Let $T_j=R^2$ before the first event of each macrostep
and $T_j=I_{2N}$ otherwise; then $\prod_jT_j=R^{2M}$.

The required local identity is
\begin{equation}
 [xy]\,\Gamma\bigl(I_{2N}+xuv^\dagger+
                      y(\Theta u)(\Theta v)^\dagger\bigr)
       =\Delta^\dagger(u)\Delta(v).
 \label{g:two-source-identity}
\end{equation}
To prove this in the occupation basis, put $U=uv^\dagger$ and
$V=(\Theta u)(\Theta v)^\dagger$. The definition of $\Gamma$ gives
$[t]\Gamma(I+tY)=d\Gamma(Y)$ for any matrix $Y$: the coefficient
replaces one creation operator at a time. Comparing the coefficient
of $xy$ in
\[
 \Gamma(I+xU+yV+xyUV)=\Gamma(I+xU)\Gamma(I+yV)
\]
therefore gives
\[
 [xy]\Gamma(I+xU+yV)=d\Gamma(U)d\Gamma(V)-d\Gamma(UV).
\]
Here the extra $xyUV$ on the left contributes $d\Gamma(UV)$ through
the linear term of $\Gamma$. Since
$UV=\langle v,\Theta u\rangle u(\Theta v)^\dagger$, the
anticommutation relations yield
\[
 \begin{aligned}
 d\Gamma(U)d\Gamma(V)-d\Gamma(UV)
  ={}&a^\dagger(u)a(v)a^\dagger(\Theta u)a(\Theta v)
       -\langle v,\Theta u\rangle a^\dagger(u)a(\Theta v)\\
  ={}&a^\dagger(u)a^\dagger(\Theta u)a(\Theta v)a(v)
   =\Delta^\dagger(u)\Delta(v).
 \end{aligned}
\]
No orthogonality between $v$ and $\Theta u$ is needed. Each source
variable occurs only linearly: in the occupation-state expansion,
two $x$ insertions contain two copies of $a^\dagger(u)$ and vanish,
and the same argument applies to $y$ and $a^\dagger(\Theta u)$.

Put $U_j=[u_j,\Theta u_j]$, $V_j=[v_j,\Theta v_j]$, and
$D_j=I_{2N}+U_j\diag(x_j,y_j)V_j^\dagger$. Second quantization gives
\[
 f(x,y)=\Tr\prod_j\Gamma(T_jD_j)
       =\det\!\left(I_{2N}+\prod_jT_jD_j\right).
\]
We linearize this determinant just as in
Section~\ref{a:cyclic-proof}, now with two source columns per event.
Let $E_j$ embed $\mathbb C^{2N}$ into block $j$ of
$\mathbb C^{2Nr}$, and set
\[
 \mathcal B_{j,j+1}=-T_{j+1}\ (j<r),\qquad
 \mathcal B_{r,1}=T_1,\qquad C=I_{2Nr}+\mathcal B,
 \qquad
 \widehat U_{:,j}=\mathcal B E_jU_j,\quad
 \widehat V_{:,j}=E_jV_j.
\]
The last two are block columns of width two. Cyclic block elimination
and the determinant lemma give
\begin{equation}
 \begin{split}
 Z_0&:=\det C=\det(I_{2N}+R^{2M})>0,
       \qquad F=\widehat V^\dagger C^{-1}\widehat U,\\
 f(x,y)&=Z_0\det\!\left(I_{2r}+F
                  \diag(x_1,y_1,\ldots,x_r,y_r)\right).
 \end{split}
 \label{g:source-determinant}
\end{equation}
The identity \eqref{g:two-source-identity} selects both sources or
neither at each physical interaction event. Consequently
\begin{equation}
 Z_{\mathrm{disc}}:=\Tr S^M
       =Z_0\sum_{J\subseteq[r]}\gamma^J\det F[J,J].
 \label{g:positive-count}
\end{equation}
Throughout this appendix, $F[A,B]$ selects whole two-dimensional
event blocks; it has size $2|A|\times2|B|$.

Each $T_j$ commutes with $\Theta$, and each frame $U_j,V_j$
intertwines the physical antiunitary with
$\Theta_1(z_1,z_2)=(-\bar z_2,\bar z_1)$. The maps in
\eqref{g:source-determinant} therefore intertwine the direct sums of
these antiunitaries. Thus $F$ is quaternionic in the event basis.
Every square block submatrix $F[A,B]$ is also quaternionic, and its
determinant is nonnegative by the Kramers determinant property
\cite[Lemma~4]{RayuduEtAl2026}. This proves nonnegativity of
\eqref{g:positive-count} without a conserved component decomposition.

\subsection{Log-concave counting and the energy estimate}

Define the mixed-minor polynomial and its complementary selector by
\begin{equation}
 \begin{split}
 g_F(y,z)&=\sum_{\substack{A,B\subseteq[r]\\|A|=|B|}}
           \det F[A,B]\,y^Az^{[r]\setminus B},\\
 q_\gamma(y,z)&=\prod_{j=1}^r(\gamma_jz_j+y_j).
 \end{split}
 \label{g:counting-polynomials}
\end{equation}
The following is the quaternionic minor-polynomial identity of
Rayudu et al.~\cite[Lemma~8]{RayuduEtAl2026}, with their variables
interchanged:
\begin{equation}
 g_F(y,z)=\sqrt{\det(D_z+F^\dagger D_yF)},\qquad
 D_y=\diag(y_1I_2,\ldots,y_rI_2).
 \label{g:quaternionic-identity}
\end{equation}
The square root is nonnegative for positive variables. Their lemma
proves that this expression is a homogeneous multiaffine polynomial
of degree $r$ with the displayed nonnegative minor coefficients.
It also proves log-concavity: $\log g_F$ is one half of the log
determinant of a positive definite matrix affine in $(y,z)$.
The polynomial $q_\gamma$ has the same properties, since it is a
product of positive linear forms. The desired coefficient is
\begin{equation}
 [y^{[r]}z^{[r]}]\,g_F(y,z)q_\gamma(y,z)
     =\sum_{J\subseteq[r]}\gamma^J\det F[J,J].
 \label{g:complementary-count}
\end{equation}
Indeed the complementary monomial of $y^Az^{[r]\setminus B}$
contains one variable per event precisely when $A=B$.

We apply the full Chen--Liu theorem, stated as
Theorem~\ref{c:chen-liu}, to \eqref{g:complementary-count}.
All input conditions are supplied exactly. A coefficient oracle
for $g_F$ checks its monomial pattern and computes a rational
determinant of order at most $2r$; the other oracle checks one
variable per pair and returns a product of rational $\gamma_j$.
The coefficients of $z^{[r]}$ in $g_F$ and $y^{[r]}$ in $q_\gamma$
are both one, giving support seeds and a normalized target at least
one. Entries of $C,\widehat U,\widehat V$ have polynomial encoding
length in $N,r,b,\log M$. The determinant and adjugate formulas
give polynomial bit bounds for $F,Z_0$. Clearing denominators and
using Hadamard's inequality, as in Section~\ref{c:oracles-section},
gives polynomial bounds on logarithms of the positive coefficient
ranges. A nonzero determinant has a nonzero Gaussian-integer
numerator after denominator clearing, so arbitrarily small positive
minors require no numerical conditioning assumption. The algorithm
uses these rational coefficient oracles, not a square-root oracle
for \eqref{g:quaternionic-identity}.

Finally set
\[
 \beta=16N/\varepsilon,\qquad
 M=\left\lceil200\beta L^2/\varepsilon\right\rceil,
 \qquad\delta=\beta/M.
\]
Estimate \eqref{g:positive-count} to relative error $1/4$ with
success probability $3/4$. If the reported value is nonpositive,
return zero; otherwise approximate its logarithm to additive error
$\beta\varepsilon/16$ and return $c$ minus that rational divided
by $\beta$. The scalar logarithm routine is
Lemma~\ref{a:scalar-logarithm}. On counting success,
\[
 |\widehat E-E_0(H)|\leq
 10\delta L^2+\frac{2N\log2}{\beta}
       +\frac{\log(4/3)}{\beta}+\frac{\varepsilon}{16}
 <\varepsilon.
\]
Here $M=O(NL^2\varepsilon^{-2})$, $r=2sM$, and the largest cyclic
matrix has order $2Nr$. Every construction, coefficient oracle,
counting call, and scalar operation has polynomial bit cost on
every execution. If $s=0$, the discrete count is the exact free
determinant $\det(I_{2N}+R^{2M})$ instead. This proves
Theorem~\ref{g:main}.

\end{document}